\documentclass[useAMS,usenatbib]{mn2e}
\usepackage{hyperref}
\usepackage{amssymb,esint}
\usepackage{graphicx}
\usepackage{epstopdf}
\usepackage{mathptmx}
\newcommand{\bea}{\begin{eqnarray}}
\newcommand{\eea}{\end{eqnarray}}
\newcommand{\beq}{\begin{equation}}
\newcommand{\eeq}{\end{equation}}

\newcommand{\Lam}{\ensuremath{\Lambda}}

\title[Self-similarity and universality of void profiles]{Self-similarity and universality of void density profiles in simulation and SDSS data}
\author[S. Nadathur et al.]{S. Nadathur$^{1}$,\thanks{seshadri.nadathur@helsinki.fi} S. Hotchkiss$^{2}$,\thanks{s.a.hotchkiss@sussex.ac.uk} J. M. Diego$^{3}$, I. T. Iliev$^{2}$, S. Gottl\"ober$^{4}$, \newauthor W. A. Watson$^{2}$ and G.~Yepes$^5$\\
$^1$Department of Physics, University of Helsinki and Helsinki Institute of Physics, P.O. Box 64, FIN-00014, University of Helsinki, Finland\\
$^2$Department of Physics and Astronomy, University of Sussex, Falmer, Brighton, BN1 9QH, UK\\
$^3$IFCA, Instituto de Fisica de Cantabria (UC-CSIC), Avda, Los Castros s/n. E-39005 Santander, Spain\\
$^4$Leibniz-Institute for Astrophysics, An der Sternwarte 16, D-14482 Potsdam, Germany\\
$^5$ Departamento de F\'isica Te\'orica, Modulo C-XI, Facultad de Ciencias, 
Universidad Aut\'onoma de Madrid, 28049 Cantoblanco, Madrid, Spain\\}

\begin{document}

\date{\today}

\pagerange{\pageref{firstpage}--\pageref{lastpage}}

\label{firstpage}

\maketitle

\begin{abstract}
The stacked density profile of cosmic voids in the galaxy distribution provides an important tool for the use of voids for precision cosmology. We study the density profiles of voids identified using the {\small ZOBOV} watershed transform algorithm in realistic mock luminous red galaxy (LRG) catalogues from the Jubilee simulation, as well as in void catalogues constructed from the SDSS LRG and Main Galaxy samples. We compare different methods for reconstructing density profiles scaled by the void radius and show that the most commonly used method based on counts in shells and simple averaging is statistically flawed as it underestimates the density in void interiors. We provide two alternative methods that do not suffer from this effect; one based on Voronoi tessellations is also easily able to account from artefacts due to finite survey boundaries and so is more suitable when comparing simulation data to observation. Using this method we show that voids in simulation are exactly \emph{self-similar}, meaning that their average rescaled profile does not depend on the void size. Within the range of our simulation we also find no redshift dependence of the profile. Comparison of the profiles obtained from simulated and real voids shows an excellent match. The profiles of real voids also show a \emph{universal} behaviour over a wide range of galaxy luminosities, number densities and redshifts. This points to a fundamental property of the voids found by the watershed algorithm, which can be exploited in future studies of voids.

\end{abstract}

\maketitle


%==============Section 1: Introduction========================%

\section{Introduction}
\label{section:intro}

Voids are large regions of space which are less dense than average and therefore show up in galaxy surveys as regions only sparsely populated by galaxies, that comprise most of the volume of the Universe. Voids are recognised as particularly interesting objects for cosmology for many reasons: they have been suggested as powerful tests the expansion history via the Alcock-Paczynski test \citep[e.g][]{Ryden:1995,Lavaux:2011yh}, their shapes, alignments and internal dynamics may be sensitive to the equation of state of dark energy (e.g. \citealt*{Park:2007,Lee:2007kq,Biswas:2010}; \citealt{Bos:2012}) or modified gravity theories \citep*[e.g.][]{Li:2009,Li:2011pj,Clampitt:2012ub}, and their abundances may be sensitive to initial conditions (\citealt*{Kamionkowski:2009}; \citealt{D'Amico:2011}). Gravitational lensing effects of voids can be measured in conjunction with other data sets \citep[e.g.][]{Krause:2013,Melchior:2013,Clampitt:2014}. It has also been suggested that the integrated Sachs-Wolfe effect of voids on the CMB can be measured to high significance \citep*{Granett:2008ju}, though this is at odds with theoretical expectations \citep*[e.g.][]{Nadathur:2011iu, Flender:2012wu}, as well as with more recent observational results \citep{Ilic:2013cn,Planck:ISW,Cai:2013ik,Hotchkiss:2014}. 

The majority of these studies rely on the assumption that voids are \emph{self-similar} objects, such that given a set of voids identified in galaxy surveys, their statistical properties do not vary across different subsets. A common manifestation of this is the assumption that the density distribution in each void can be simply rescaled depending on the size of the void, and that following such a rescaling the average void density profile no longer depends on the void size. This assumption is particularly important for those studies \citep[e.g.][]{Lavaux:2011yh,Sutter:2012tf,Krause:2013,Melchior:2013} which depend on an explicit form for the mean density profile, but is also present in implicit form in many other analyses.  A related but distinct assumption sometimes employed is that such a rescaling is \emph{universal}---that is, that the rescaled void properties are independent of the properties of the tracer population in which the voids were identified or the survey redshift.

In fact the form of the mean void density profile has also been a subject of much study (e.g. \citealt{Colberg:2005,Lavaux:2011yh,Pan:2011hx,Ceccarelli:2013,Nadathur:2014a,Sutter:2013ssy}; \citealt*{Ricciardelli:2014,Hamaus:2014fma}) but there is no consensus on the functional form of such a profile, either from simulation or galaxy data. This is in large part due to a similar lack of consensus on how to define a `void': there are a large number of different void-finding algorithms in use, which do not always return precisely the same set of objects \citep{Colberg:2008}. However, there are also different techniques for measuring the average profile of a given set of voids. Most involve counting numbers of all tracer particles (either galaxies or dark matter particles in simulations) contained with regions of a known volume and normalizing with respect to the mean, but different weighting schemes --- of different statistical merit --- may be employed in the subsequent averaging process. There is also no firm consensus yet on whether profiles in simulation match those seen in real galaxy data, partly because of different definitions of a void applied in the two cases, but probably also because the profile measurement technique requires refinement in the presence of complex survey masks and holes. 
 
Some recent attempts have been made to assess the assumptions of self-similarity and universality, but here again the results appear in conflict with each other. Preliminary work \citep{Nadathur:2014a} suggested that voids in SDSS data do show some degree of universality of profiles, but did not examine self-similarity. \citet{Ricciardelli:2014} find that voids in simulation are exactly self-similar (no dependence of the average profile on void size), but do find some dependence on the tracer galaxy luminosity. On the other hand, \citet{Hamaus:2014fma,Sutter:2013ssy} find almost the opposite result: that the density profile strongly depends on the size of the voids in the sample, but that the form of this size dependence is roughly unchanged by differences in the tracer population.

In this paper we use samples of voids found in realistic mock LRG catalogues from the Jubilee $N$-body simulation \citep{Watson:2013mea} and in data from SDSS galaxy catalogues to investigate these issues. The voids are identified using a modified version of the {\small ZOBOV} watershed transform void finder \citep{Neyrinck:2007gy} and are chosen according to strict criteria in order to avoid spurious detections arising from Poisson noise or survey boundary contamination effects. We discuss the question of robust measurement of the density profile in the presence of Poisson noise and with the added complication of survey boundaries as encountered in real galaxy survey data. We show that the simplest profile estimation method, as used by \citet{Nadathur:2014a,Hamaus:2014fma} among others, is statistically poorly formulated and systematically underestimates the density in the low-density interiors of voids due to Poisson noise effects. The degree of this bias is also unfortunately dependent on the void size, being greater for smaller voids. Applying a volume-weighted averaging procedure corrects for this effect, but both methods still suffer from artefacts due to the finite nature of the surveyed region. This last effect is not important in simulations if the entire simulation box is used, but is certainly significant for application to real galaxy survey data. 
     
We therefore propose a new method of measuring the density profile of voids based on a reconstruction of the density field using Voronoi tessellations. Tessellation field estimators have long been recognised as a method of reconstructing the density field from a discrete distribution of point objects that minimize Poisson noise \citep{Schaap:2007,vdWSchaap:2008}, and this is in fact already the method used by the {\small ZOBOV} void-finder to find density minima in the tracer distribution. We demonstrate that an additional advantage of this method is that it can easily account for the survey geometry and thus should be preferred when comparison between simulation and observation is desired.

Using this profile estimator we examine the assumption of self-similarity of voids in our simulation. We find that our void selection criteria, which are independently motivated by statistical considerations, produce a sample of voids that \emph{is} self-similar. That is, the stacked density profile shows no dependence on void size, and is also independent of redshift within the extent of our simulation. 

We then compare profiles of our simulated voids with those obtained from voids in the SDSS data, and find an excellent match between the corresponding LRG samples. Even more interestingly, the profiles of voids from all the real galaxy catalogues --- spanning both Main Galaxy and LRG catalogues, a wide range of different luminosities and three orders of magnitude in number density --- agree very well with each other. This universality points to a fundamental property of the voids obtained by our algorithm, and suggests that precision measurements of the void density profile will be an important tool in future cosmological studies.

The layout of the paper is as follows. In Section~\ref{section:simulation} we briefly describe the Jubilee simulation, the creation of the mock LRG catalogues and the void-finding procedure. The data catalogues from SDSS are presented briefly in Section~\ref{section:catalogue}. In Section~\ref{section:profiles} we describe three different methods for estimating the mean profile of a given stack of voids and compare their relative merits and failings. Sections~\ref{section:self-similarity} and \ref{section:universality} present the main results of our paper, which is that our independently motivated selection criteria for voids produce a sample of objects that are exactly self-similar, with density profiles that do not depend on the void size or redshift; and that density profiles from SDSS data not only agree with this but also demonstrate a degree of universality. Our results differ somewhat from the conclusions reached by previous groups; in Section~\ref{section:previous_results} we discuss possible reasons for this difference. Finally we summarize the implications of our findings in Section~\ref{section:conclusions}.

%==============Section 2: Simulation========================%

\section{Voids in simulation}
\label{section:simulation}

Our simulation results are based on the Jubilee $N$-body simulation \citep{Watson:2013mea}, which has $6000^3$ particles in a box volume of $(6\;h^{-1}\rmn{Gpc})^3$. This corresponds to an individual particle mass of $7.49\times10^{10}\;h^{-1}\rmn{M}_\odot$ and a minimum resolved halo mass (with 20 particles) of $1.49\times10^{12}\;h^{-1}\rmn{M}_\odot$. The simulation uses cosmological parameters based on the Wilkinson Microwave Anisotropy Probe (WMAP) 5-year results when combined with external data sets \citep{Komatsu:2008hk}, i.e. $\Omega_\rmn{m}=0.27$, $\Omega_\Lam=0.73$, $h=0.7$, $\Omega_\rmn{b}=0.045$, $\sigma_8=0.8$ and $n_\rmn{s}=0.96$. The initial conditions for the simulation were set using the Zel'dovich approximation \citep{Zeldovich:1969sb} at a redshift of $z=100$ and evolved forward \citep[see][]{Watson:2012mt}.

Note that the size and halo mass resolution of the Jubilee simulation allows for the construction of full-sky mock catalogues even at high redshifts without repetition of the simulation box. Indeed the Jubilee box is complete for a central observer out to a redshift of $z\sim1.4$.

%==========Subsection 2.1: LRGs=============%
\subsection{Mock LRG catalogues}
\label{subsection:LRGs}

To model the LRG population in our survey we use a Halo Occupation Distribution (HOD) model on the Jubilee simulation. We first use an on-the-fly halo-finder based on the Spherical Overdensity (SO) algorithm \citep{Lacey:1994su} to resolve galaxy size haloes along the light cone down to a mass of $\sim10^{12}\;h^{-1}\rmn{M}_\odot$ and then populate these haloes with LRGs according to the HOD model according to the results of \citet{Zheng:2008np}. This model is calibrated on a sample of SDSS LRGs with $g$-band magnitudes $M_g<-21.2$ between redshifts of $0.16$ and $0.44$ \citep{Eisenstein:2005su}. Such LRGs typically reside in haloes of mass in excess of $\sim10^{13}\;h^{-1}\rmn{M}_\odot$ \citep{Zheng:2008np,Wen:2012tm,Zitrin:2012}, which is well above the resolution limit of Jubilee. In applying the best-fit model parameters from \citet{Zheng:2008np}, we assume that the LRGs are the brightest cluster galaxy (BCG) in their halo. This means we have ignored the small fraction ($\sim5\%$) of LRGs that are satellite galaxies.

We assign luminosities to the LRG population based on the host halo masses \citep{Zheng:2008np}, taking into account the varying steepness of the mass-luminosity relationship as a function of halo mass \citep[see][for details]{Watson:2013cxa}. We then apply a log-normally distributed random scatter between the LRG location and that of the dark matter density peak to match the results of \citet{Zitrin:2012} for BCGs. Finally we assign the host halo bulk velocity to the LRG and include this as a Doppler correction term to the `observed' redshift of the LRGs. To mimic the effect of peculiar velocities in smoothing out or distorting voids in the real data, we convert these `observed' redshifts into `observed' LRG positions in comoving coordinates using our fiducial cosmology. For further details and discussion of all of these modelling steps we refer the reader to the discussion in \citet{Watson:2013cxa}.

To this population of mock LRGs we now apply magnitude and redshift cuts to construct two mock full-sky LRG samples. We select mock LRGs with $-23.2<M_g<-21.2$ and redshift $0.16<z<0.36$ to create the `Jubilee Dim' (JDim) sample, and those with $-23.2<M_g<-21.8$ and $0.16<z<0.44$ to create the `Jubilee Bright' (JBright) sample as in \citet{Hotchkiss:2014}. These mock samples are intended to match the properties of the actual (quasi-) volume-limited SDSS `Dim' and `Bright' LRG samples from \citet{Kazin:2010}, from which some of the catalogues of voids presented in \citet{Nadathur:2014a} were drawn. 

%=========Subsection 2.2: Void-finding===========%
\subsection{Void-finding}
\label{subsection:void-finding}

To identify voids in the mock LRG catalogues we use a modified version of the watershed void-finder {\small ZOBOV} \citep{Neyrinck:2007gy} following closely the procedure outlined in \citet{Nadathur:2014a}. 

The principle upon which {\small ZOBOV} is based is the reconstruction of the density field of a discrete distribution of tracers (in our case LRGs) through the use of a Voronoi tessellation field estimator or VTFE \citep{Schaap:2007,vdWSchaap:2008}. Each particle location is assigned a density inversely proportional to the volume of that particle's Voronoi cell. Tessellation-based density estimators have been used as the starting point of many different void-finding techniques, e.g. \citet{Platen:2007qk,Aragon-Calvo:2010,Platen:2011,Sousbie:2011a}. The tessellation scale is fully-self adaptive and thus gives a local estimate of the density at the location of the individual tracer particle that is much less prone to annoying shot noise effects that plague grid-based methods \citep{Neyrinck:2007gy}, without relying on arbitrary smoothing kernels. It also offers a natural way to correct for local variations in the mean density due to selection effects \citep{Nadathur:2014a}, which is important for some of our SDSS samples. 

As {\small ZOBOV} is designed to operate on a cubic box whereas our distribution of mock LRGs occupy an annular shell around the observer, we follow the usual procedure of enclosing the mock galaxies within a buffer of boundary particles at both the lower and upper redshift caps. It is important that these boundary particles are sufficiently densely packed to ensure that no Voronoi cells leak outside the survey volume and that `edge' LRGs adjacent to boundary particles are identified and appropriately handled; our method here exactly follows that outlined in detail in \citet{Nadathur:2014a}. 

Starting with the VTFE reconstructed density field following this step, we then identify local minima or zones and merges these zones together to form voids, as described by \citet{Neyrinck:2007gy}. This procedure is in principle parameter-free; however, in practice it is necessary to introduce some parameters to halt the merging of zones and obtain a usable catalogue of independent, non-overlapping voids that are statistically distinct from spurious detections in random point distributions \citep{Nadathur:2014a}.

To do this we introduce the following general criteria. To qualify as a starting seed, we require that a zone must have a minimum density $\rho_\rmn{min}$ below a specified threshold value; this serves to remove spurious shot noise fluctuations and local density minima within large-scale overdensities, which do not qualify as voids. Then, zones neighbouring a deeper qualifying seed zone are successively merged with their deepest neighbour provided that (a) the watershed link density $\rho_\rmn{link}$, and (b) the \emph{density ratio} $r$ between the link density and the minimum density of the zone(s) being added, are both below specified thresholds. The first of these conditions prevents the voids from growing to include regions of high density, and the second ensures that statistically significant subvoids \citep{Neyrinck:2007gy} are treated as independent rather than grouped together.

It is important to recognize that the specific threshold values chosen for each of these quantities can be physically motivated but remain to some extent arbitrary. The most appropriate choice may depend on the purpose for which the voids are to be studied. \citet{Nadathur:2014a} provided two possible choices of thresholds: for `Type1' voids, $\rho_\rmn{min}<0.3\overline{\rho}$ (where $\overline{\rho}$ is the mean density of the sample), $\rho_\rmn{link}<\overline{\rho}$ and $r<2$, while `Type2' voids were defined by $\rho_\rmn{min}<0.2\overline{\rho}$, $\rho_\rmn{link}<0.2\overline{\rho}$ and $r<2$. The cut on $\rho_\rmn{min}$ means that Type1 voids are statistically distinct from fluctuations due to Poisson noise at the $3\sigma$-equivalent confidence level, but are otherwise minimally defined. The Type2 definition is much stricter, but consequently returns many fewer voids. 

In this paper we will introduce a third class of voids: `Type3' voids are those defined by $\rho_\rmn{min}<0.3\overline{\rho}$, $\rho_\rmn{link}<0.3\overline{\rho}$ and $r<2$. Thus Type3 voids retain the statistical significance of Type1, while the stricter merging criteria ensure that the interiors of voids are less contaminated by intervening high-density ridges and filaments, and that the number of voids obtained is larger (since large composite voids are more likely to be split into independent constituents). In total, we obtain 1134 such voids from the JDim sample, and 769 from the JBright sample. From these---unless otherwise stated---we further exclude those outlier voids composed of 5 or more merged zones, thus reducing the sample sizes of Type3 voids to 1087 and 748, respectively. The reason for this is explained in more detail in Section~\ref{section:self-similarity}. Unless otherwise stated, all simulation results in this paper are illustrated with Type3 voids from JDim.

Within this sample, we identify `edge' voids as those to which `edge' galaxies would have been assigned had we not removed them. In other words, all edge voids contain a mock galaxy which is adjacent (in a Voronoi sense) to a galaxy which is itself adjacent to one of the boundary particles in the buffer (such edge galaxies are never themselves included as members of a void). This is a conservative classification. Edge voids may possibly have been artificially truncated by boundary effects. Due to the full-sky nature of the Jubilee simulation, such edge voids are naturally concentrated near the two redshift caps of the survey.

For each void, we define the void centre to lie at the Voronoi volume-weighted barycentre of its member galaxies, $\mathbf{X}=\frac{1}{\sum_i V_i}\sum_i \mathbf{x}_i V_i$, and the void radius $R_v$ to be the volume of a sphere occupying a volume equal to the sum of the Voronoi volumes of the void member galaxies.

%==============Section 3: Catalogue========================%

\section{Voids in SDSS galaxy surveys}
\label{section:catalogue}

\begin{table}
\begin{centering}
\caption{Numbers of voids in different galaxy samples}
\begin{tabular}{@{}ccrrr}
\hline
Sample name & Sample type & \multicolumn{3}{c}{Number of voids} \\
 &  & Type1 & Type2 & Type3 \\
\hline
 \emph{dim1} & SDSS Main & 80 & 53 & 85 \\
\emph{dim2} & SDSS Main& 271 & 199 & 281 \\
\emph{bright1} & SDSS Main & 262 & 163 & 300 \\
\emph{bright2} & SDSS Main & 112 & 70 & 149 \\
\emph{lrgdim} & SDSS LRG & 70 & 19 & 147 \\
\emph{lrgbright} & SDSS LRG & 13 & 1 & 27\\
JDim & mock LRG & 657 & 377 & 1087\\
JBright & mock LRG & 342 & 166 & 748\\
\hline
\end{tabular}
\label{table:numbers}
\end{centering}
\end{table}

The identification of voids in the SDSS DR7 galaxy samples has been discussed in detail in our earlier work \citep{Nadathur:2014a}, which provided a public catalogue of voids.\footnote{This catalogue is available for download from \url{http://research.hip.fi/user/nadathur/download/dr7catalogue/}. Although catalogues of Type3 voids were not originally provided, they can be extracted easily from the available data using the software tools provided with the download.} 

This catalogue contains voids identified in six different spectroscopic volume-limited or quasi-volume-limited samples \emph{dim1}, \emph{dim2}, \emph{bright1}, \emph{bright2}, \emph{lrgdim} and \emph{lrgbright} drawn from DR7 redshift surveys, both from the main galaxy and LRG catalogues. The first four samples are subsets of the main galaxy sample of the New York University Value-Added Galaxy Catalog \citep[NYU-VAGC][]{Blanton:2004aa}, which is a catalogue of low-redshift ($z\lesssim0.3$) galaxies based on publicly-released surveys matched to galaxies from the SDSS \citep{Abazajian:2008wr} using improved photometric calibrations \citep{Padmanabhan:2007zd}. The (evolution- and $K$-corrected) $r$-band magnitude and redshift limits of the samples are \emph{dim1}: $M_r<-18.9$ and $0<z<0.05$; \emph{dim2}: $M_r<-20.4$ and $0<z<0.1$; \emph{bright1}: $M_r<-21.35$ and $0<z<0.15$; \emph{bright2}: $M_r<-22.05$ and $0<z<0.2$. The two LRG samples are taken from \citet{Kazin:2010}: the $g$-band magnitude limits and redshift extents match those of the Dim and Bright samples in that work but also include galaxies from the stripes in the southern Galactic hemisphere. The \emph{lrgdim} and \emph{lrgbright} samples are therefore directly comparable to the JDim and JBright simulation samples described in Section~\ref{subsection:LRGs}.

From these galaxy samples we extract voids matching the type definitions described in Section~\ref{subsection:void-finding}. Compared to the original catalogues presented in \cite{Nadathur:2014a} the Type3 definition introduced above provides a cleaner sample than the Type1 category, and a larger number of voids than either Type1 or Type2. The numbers of structures obtained are summarised in Table~\ref{table:numbers}.

Our void type definitions and treatment of survey boundaries and the redshift-dependent variation of the mean galaxy density mean that our catalogues are free of spurious contamination due to shot noise, survey boundaries and selection effects which have plagued some previous catalogues \citep[see the discussion in][]{Nadathur:2014a}. We also emphasise that our mock LRG catalogues are constructed on the light cone, and that we employ \emph{exactly} the same algorithm for identification of voids in the simulated LRG catalogues from Jubilee as in the real data. This allows a direct comparison between simulation and data.

Nevertheless, the SDSS survey geometry and bright star mask may introduce effects which are not present in the simulation. In particular, the complex boundary means that the vast majority of real voids are `edge' voids in the conservative sense defined in Section~\ref{subsection:void-finding}. Voids in simulation are therefore representative of a slightly different population of objects than those in the real data. We discuss the possible effects of this in Section~\ref{section:universality} when we compare simulation to observation. These effects could be studied in more detail by applying the SDSS mask to simulation data as well, but we leave this for future work.

%==============Section 4: Profiles ==========================%

\section{Measuring void density profiles}
\label{section:profiles}

In this Section we introduce and discuss three different methods of measuring the average profile of a stack of voids based on the distribution of the tracer particles used to identify them. The first and simplest of these methods, which has been commonly used \citep[e.g.][]{Sutter:2013ssy,Nadathur:2014a,Hamaus:2014fma} is based simply on counting the number of tracer particles within regions of known volume and averaging the result over the stack of voids. We shall refer to this method as the `naive method' and show that it is systematically biased low in underdense regions near the centres of voids. In addition, this bias is worse for smaller voids, which may introduce a spurious dependence of reconstructed densities on void size. We propose two alternatives: one is an averaging scheme based on volume-weighting which reduces these effects, and which we shall refer to as the `Poisson method', for reasons which will shortly be made clear. The other is a true estimate of the VTFE reconstructed density field seen by {\small ZOBOV}, which we refer to as the VTFE method.

%==================Fig. 1: naivefailings=======================%
\begin{figure*}
%\begin{center}
\includegraphics[width=80mm]{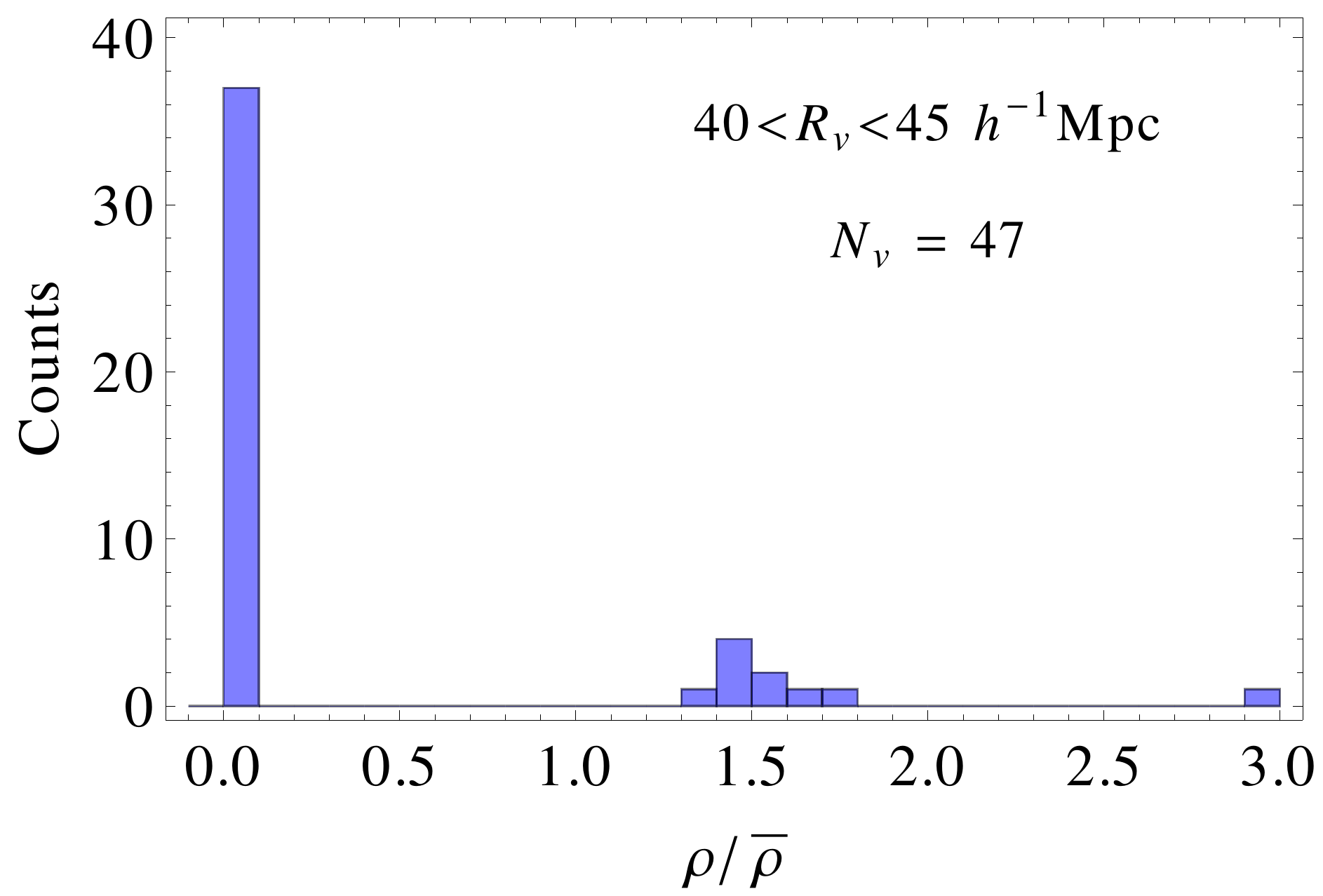}\hspace{3em}\includegraphics[width=77mm]{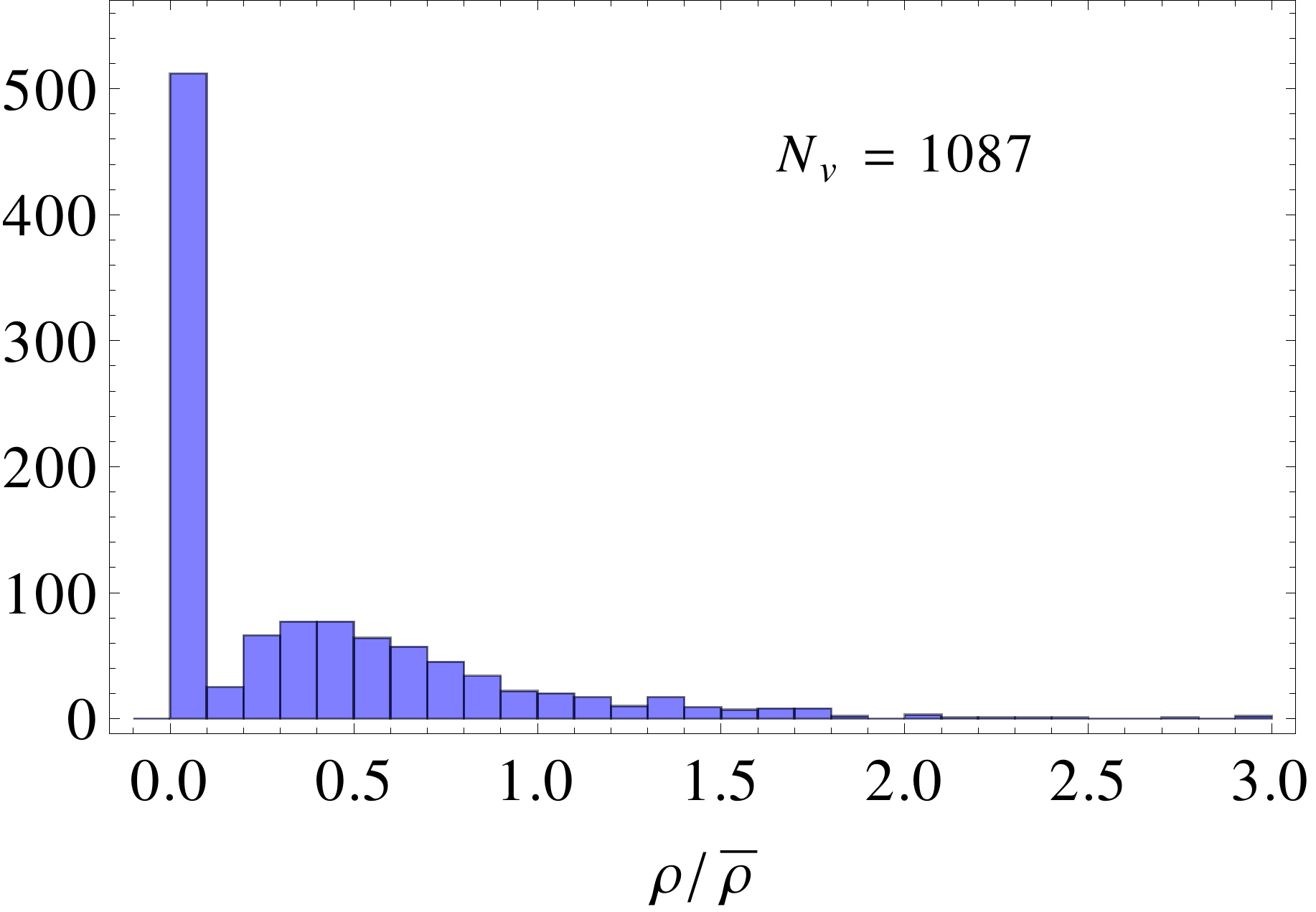}
\caption{Histogram of the density measurements as reconstructed by equation~\ref{eq:naive1} in a single interior shell (centred at rescaled distance $r=0.2R_v$ from the void centres) for Type3 voids from the JDim mock LRG sample. \emph{Left:} Using only the 47 voids with radii between 40 and 45 $h^{-1}$Mpc. The distribution is clearly multimodal, with the vast majority of shells having no tracer galaxies at all, thus returning an unphysical reconstructed density of zero. \emph{Right:} The same as the left panel, but including all 1087 voids. Due to the range of shell volumes the Poisson nature of the distribution visible in the left panel gets smeared out, but the predominant mode at zero density remains.} 
\label{figure:naivefailings}
%\end{center}
\end{figure*}
%==================Fig. 1: naivefailings=======================%

As is usually the case in profile measurements, all three methods actually measure the local \emph{number} density of the tracer population rather than mass density. This is necessary when dealing with voids in real galaxy distributions where the galaxy masses may not be known accurately, and it is therefore preferable to do the same with the Jubilee simulation. Note that it is also the number density rather than the mass density that is used by {\small ZOBOV}. In cases where tracer particle masses are known simple extensions of our methods can provide mass density measurements too. In what follows we will use the symbol $\rho$ to refer to the number density.

%=========Subsection 4.1: Naive method===========%
\subsection{Naive method}
\label{subsection:naive}

In this method the density profile of the $i$th void is obtained by measuring the number density of tracer particles (galaxies) in spherical shells of a given width and different radial distances from the void centre. Self-similarity of the voids is assumed in order to recale all distances in units of the void radius $R_v^i$ and the set of rescaled profiles is then stacked and averaged in each bin of radial values $r/R_v$. Formally, the measured density of the $i$th void in the $j$th radial shell is 
\beq
\label{eq:naive1}
\rho_i^j = \frac{N_i^j}{V_i^j}\,,
\eeq
where $N_i^j$ is the total number of tracer particles within shell volume $V_i^j$ at scaled radial distance $r_j/R_v^i$ from the void centre. We do not restrict the particle counts $N_i^j$ to only those identified as members of the given void by the {\small ZOBOV} algorithm but include all tracer particles that fall within the shell. Shells are taken to have a standard radial thickness $\Delta$ in units of the void radius so that varying $\Delta$ determines the resolution with which the profile can be measured. The average density of the stack of voids in the $j$th shell is then simply taken to be
\beq
\label{eq:naive2}
\overline{\rho}^j = \frac{1}{N_v}\sum_{i=1}^{N_v} \rho_i^j\,,
\eeq
and errors in the mean are estimated from the standard deviation:
\beq
\label{eq:naive3}
\Delta\rho^j=\frac{1}{\sqrt{N_v-1}}\sigma_{\rho}\,,
\eeq
where $\sigma_\rho$ is the standard deviation of $\rho_i^j$ values. 

However, there are several problems with this approach. The first and most obvious is that since \emph{individual} voids are highly elliptical and the survey geometry is in general quite complex, spherically symmetric shells around the void centre will often partially extend beyond the surveyed region, where the tracer population is by definition absent. Unless this effect is corrected for each time, such shells will give artificially low density values, particularly at larger radial distances. We will show in Section~\ref{subsection:comparison} that this already becomes important at radial distances $r\sim R_v$. It is in principle possible---though for realistic survey masks, computationally expensive---to correct for this leakage effect. However, such a correction has not previously been implemented in the literature.

A less avoidable objection to the naive method is that it is statistically biased. It is clear that under the assumption of self-similarity, the shell occupancies $N_i^j$ are random variables drawn from Poisson distributions with means $\overline{\rho}^jV_i^j$. This results in a distribution of $\rho_i^j$ values from equation~\ref{eq:naive1} as shown in Fig.~\ref{figure:naivefailings}. For a stack containing only voids with similar radii $R_v$, the distribution of $\rho_i^j$ values is close to bimodal with one mode at zero density. Broadening the range of void radii obscures this Poisson nature somewhat because of the range of shell volumes $V_v^j$, but the predominance of the mode at $\rho_v^j=0$ remains clear. However, the physical density seen by {\small  ZOBOV} is in fact nowhere zero: zero number counts are resolution artefacts due to shell volumes that are too small. 

More formally, a property of the Poisson distribution is that, assuming a uniform prior on the underlying expectation value $\lambda$,
\beq
\label{eq:PoissonPDF}
P(\lambda|N)=P(N|\lambda)=\frac{\lambda^N\exp(-\lambda)}{N!}\,,
\eeq
from which one can deduce that, given an observation of $N$ tracer particles, the expectation value of $\lambda$ is in fact
\beq
\label{eq:PoissonExp}
E\left[\lambda|N\right] = \int \lambda\,P(\lambda|N)\,\rmn{d}\lambda = N+1>N\,.
\eeq
Comparison of equations~\ref{eq:naive1} and \ref{eq:PoissonExp} shows that if the tracer population Poisson samples an underlying number density value in each shell, the naive method of reconstruction \emph{systematically underestimates} this density when $N$ is small. This discreteness bias is important in the low-density void interiors. It is also worse for smaller shell spacing $\Delta$ (i.e. higher resolution measurement of the profile), and for smaller voids (as the shell volumes are fixed as a proportion of void volumes).

\begin{figure}
%\begin{center}
\includegraphics[width=85mm]{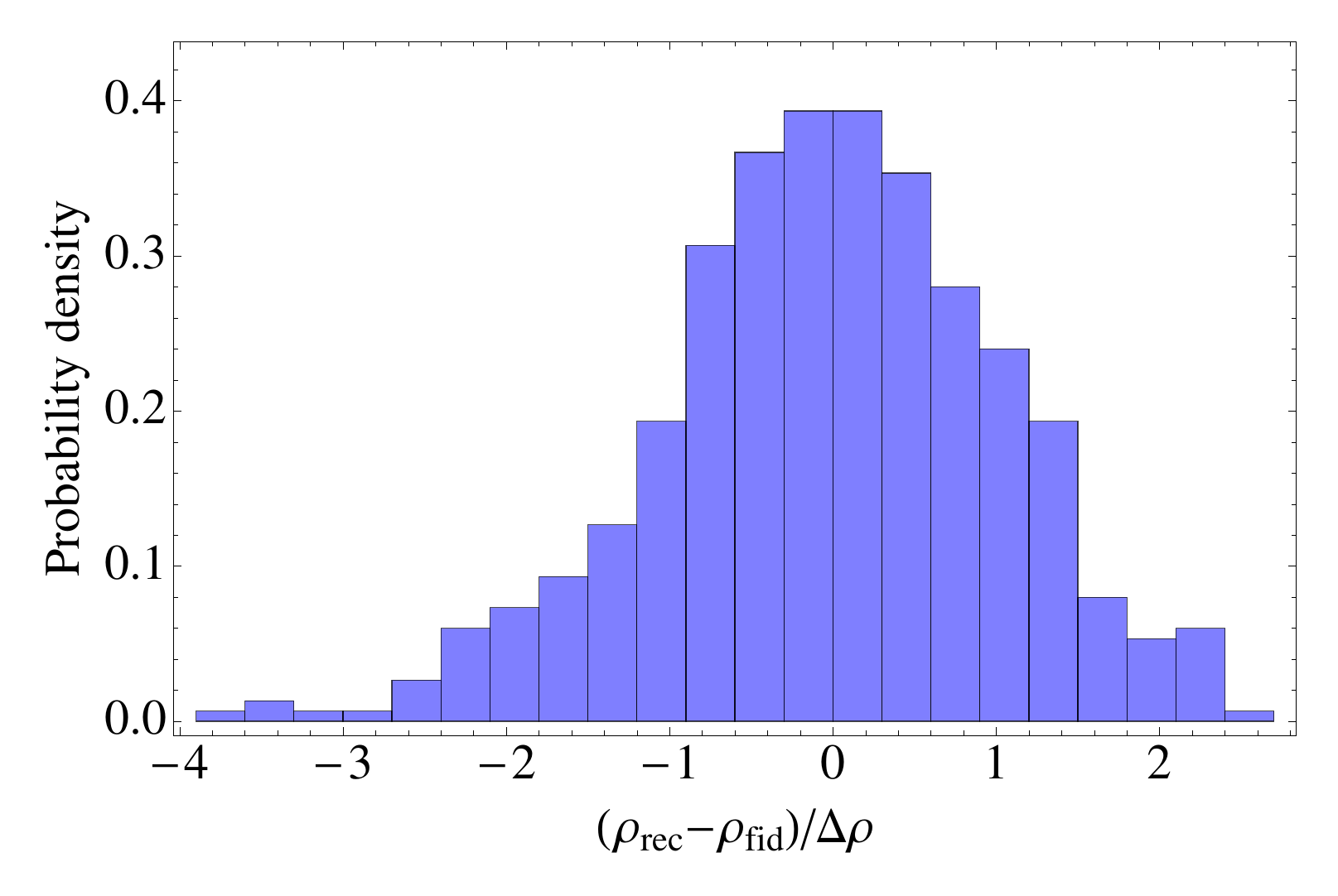}
\caption{Probability density histogram for the difference between the fiducial density and the mean density reconstructed using equation~\ref{eq:naive2}, in units of the estimated error in the mean from equation~\ref{eq:naive3}. The number of shells were chosen to match typical void numbers found in real void catalogues, and their volumes were chosen to match the distribution of volumes seen for an interior shell in voids from our JDim sample. The shells were populated with tracer particles by Poisson sampling the fiducial density. There is a clear skew towards $\rho_\rmn{rec}<\rho_\rmn{fid}$, and large deviations are more common than would be expected for a Gaussian-distributed error in the mean.} 
\label{figure:naivebias}
%\end{center}
\end{figure}

Related to this is an issue regarding the error estimation. Since the distribution of $\rho_i^j$ values shown in Fig.~\ref{figure:naivefailings} is very far from Gaussian, the error in the reconstructed mean can be larger than a naive interpretation of equation~\ref{eq:naive3} would indicate. To demonstrate this we ran a simple simulation in which regions with a range of different volumes were populated with tracer particles by Poisson sampling a single fiducial mean density $\rho_\rmn{fid}$, and the average density and estimated error were then reconstructed according to equations~\ref{eq:naive1} and \ref{eq:naive2}. This process was repeated 500 times. Fig.~\ref{figure:naivebias} shows the resultant probability density histogram for the difference between reconstructed and fiducial densities in units of the estimated error. There is a clear skew towards negative values. In addition it can be seen that large negative discrepancies are not uncommon: even for just 500 simulated reconstructions, values of $|\rho_\rmn{rec}-\rho_\rmn{fid}|\sim4\Delta\rho$ do occur, demonstrating the problem with interpreting $\Delta\rho$ in terms of confidence levels. 

%==================Fig. 2: naive bias=======================%
%==================Fig. 2: naive bias=======================%

We conclude that although this method of density reconstruction has been commonly used in the past \citep[e.g.][]{Nadathur:2014a,Sutter:2013ssy,Hamaus:2014fma}, it is statistically flawed and therefore unsuitable for measurement of void density profiles. We do not consider it further.

%=========Subsection 4.2: Poisson method===========%
\subsection{Poisson method}
\label{subsection:Poisson}

Since each of the $N_i^j$ values is a Poisson distributed random variable, their sum will also be Poisson distributed. Therefore the correction that should be made to the naive method is to redefine the average density in the $j$th radial bin as
\beq
\label{eq:Poisson1}
\overline{\rho}^j = \frac{\left(\sum_{i=1}^{N_v} N_i^j\right)+1}{\sum_{i=1}^{N_v} V_i^j}\,,
\eeq 
where the sums run over all $N_v$ voids in the stack. Note that this is equivalent to replacing the simple averaging of equation~\ref{eq:naive2} by a \emph{volume-weighted} average. The resulting expression is simply the total number of particles in the $j$th shell of all voids, divided by the total volume of those shells (with the additional $+1$ added to correct for the systematic bias discussed in equation~\ref{eq:PoissonExp}). It is therefore by definition the true average number density in that shell. This is a strong argument in favour of always using a volume weighting when calculating average densities.

Since the sum $\sum_{i=1}^{N_v} N_i^j$ is a Poisson-distributed random variable, we can estimate the error in $\overline{\rho}^j$ defined above at any required confidence level directly from the probability distribution equation~\ref{eq:PoissonPDF}. In this paper we choose the $68\%$ confidence level to define the error estimate. Note that in general this will give \emph{asymmetric} error bars when number counts are small in the interiors of voids.

Not only is this method better motivated from a formal statistical viewpoint, as a by-product it also sharply reduces the noise in measurements $\overline{\rho}^j$ in the void interior. This is because it down-weights the contributions from the smallest shells, which are most susceptible to discreteness noise, relative to those from the larger shells where this effect is less important, as $N_i^j$ in larger. This is a further justification in favour of volume-weighted averaging. An alternative procedure of weighted averaging has been used by \citet{Ricciardelli:2014}, although in that case weights are chosen explicitly to down-weight the effect of outliers, rather than based on a physical justification. 

However, while this method of profile measurement is a significant improvement on the naive method, it still has some of the same disadvantages. The most important of these is that shells may still extend beyond survey boundaries, and unless this volume leakage effect is explicitly corrected for, artefacts are introduced in the profile at large distances from the void centre. Another problem is that correction for the redshift-dependence of the local mean tracer density is difficult. Therefore we now describe a method based the VTFE density reconstruction that addresses these issues.

%=========Subsection 4.3: VTFE method===========%
\subsection{VTFE method}
\label{subsection:VTFE}

Both of the methods of profile measurement described so far discard the information about the VTFE reconstructed density field actually used by the void-finding algorithm in favour of cruder counts-in-shells estimates, which are more prone to Poisson noise. As we have already argued, the self-adapative nature of the tessellation estimator and the large number of point positions contributing to each Voronoi cell make this our best estimate of the local density at each particle location. This is why {\small ZOBOV} uses the VTFE method in the first place. 

We therefore propose the following estimator for the stacked density in the $j$th radial shell from the void centre, which makes use of the VTFE reconstructed density information:
\beq
\label{eq:VTFE1}
\overline{\rho}^j = \frac{\sum_{i=1}^{N_v}\sum_{k=1}^{N_i^j} \rho_k V_k}{\sum_{i=1}^{N_v}\sum_{k=1}^{N_i^j} V_k}\,,
\eeq
where $V_k$ is the volume of the Voronoi cell of the particle $k$, $\rho_k$ is its density inferred from the inverse of the Voronoi volume \citep[with additional corrections applied for the redshift-dependence of the local mean where necessary, see][]{Nadathur:2014a}; the sum over $k$ runs over all particles in the $j$th shell of void $i$ (not only void member particles); and the sum over $i$ includes all voids in the stack. This is again in effect a volume-weighted average of the density values at each particle in the stack, the difference compared to the Poisson method being that the density estimate is taken directly from the VTFE, rather than from cruder number counts.

To estimate the error in $\overline{\rho}^j$ determined in this way we construct $N_v$ jacknife samples $\overline{\rho}_J^j$, $J=1,\ldots N_v$, of the density measurement by excluding all particles from each of the $N_v$ voids in turn, and estimate the error in the mean due to variation in individual voids as 
\beq
\label{eq:VTFE2}
\Delta\rho^j = \sqrt{\sum_{J=1}^{N_v}\left(\overline\rho_J^j-\frac{1}{N_v}\sum_{K=1}^{N_v}\overline\rho_K^j\right)^2}\,.
\eeq
Note that this represents the degree of variation in density values across the sample of voids, and is therefore conceptually different to the error estimation procedure in Section~\ref{subsection:Poisson}, which captured instead the measurement uncertainty in the overall mean using the Poisson method. 

When a particular void has no representative particles within a given shell, as occurs often in the smallest interior shells (see Fig.~\ref{figure:naivefailings}), rather than assign an unphysical zero density to that shell we extrapolate the measured density value from the nearest filled shell for that void. In practice this involves `copying' particles from the nearest occupied shell and evaluating the sum over $k$ for these particles. In effect this means that we are appropriately adjusting the spatial resolution of the measurement based on the size of the void. A less optimal way of handling this situation would be to simply exclude that void entirely from the density determination in that shell, but this would lead to the measured value being significantly biased high, since voids with higher densities at a particular radial distance are more likely to have particles in that shell.

%==================Fig. 3: types=======================%
\begin{figure}
%\begin{center}
\includegraphics[width=85mm]{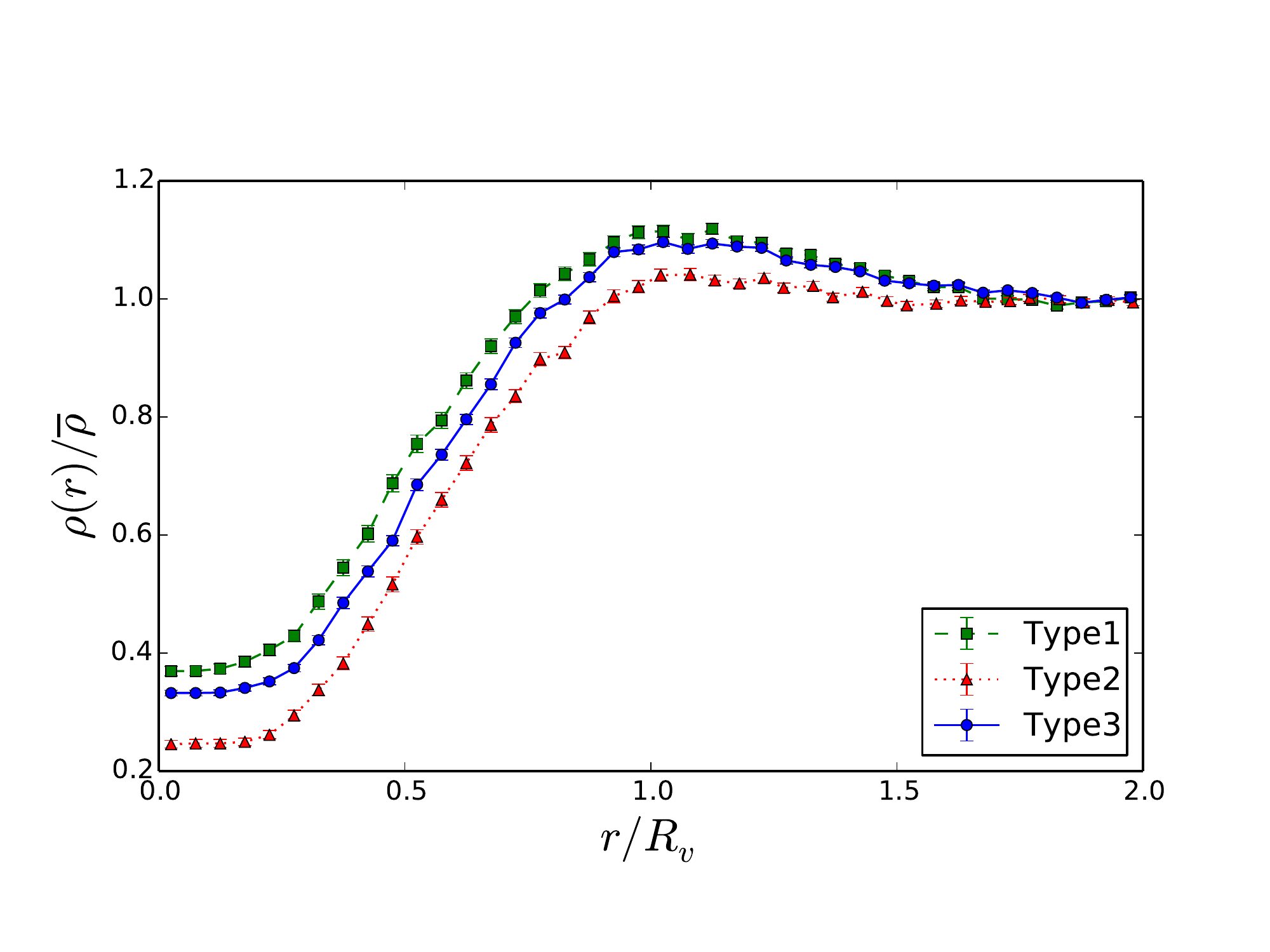}
\caption{Stacked density profiles for Type1, Type2 and Type3 voids from the JDim mock LRG sample, as reconstructed using the VTFE method described in Section~\ref{subsection:VTFE}. The difference in the stacked profile caused by the selection cut on $\rho_\rmn{min}$ can be seen by comparing types 1 and 3 with Type2. The smaller difference between Type1 and Type3 is due to the stricter controls on zone merging applied in the latter case.} 
\label{figure:types}
%\end{center}
\end{figure}
%==================Fig. 3: types=======================%

This method of dealing with empty shells means that profile measurements $\overline{\rho}^j$ in neighbouring radial bins $j$ are not independent of each other, and the degree of correlation increases if the bin width $\Delta$ is decreased. This should be borne in mind when interpreting the errors calculated from equation~\ref{eq:VTFE2}! In practice this correlation is important out to bins at radial distances of $\sim0.5R_v$, after which all voids have representative particles in every bin and the extrapolation procedure is not required. 

The use of the Voronoi tessellation means that density and volume measurements are naturally tied to particle locations. Therefore as long as the treatment of boundary effects is sufficiently robust to prevent the Voronoi cells themselves from leaking out of the survey volume \citep[as is necessary for a consistent application of the void-finding algorithm; see][for a detailed discussion]{NH:2013b,Nadathur:2014a} shells partially extending outside the survey volume do not pose a problem for the VTFE method no matter how complex the survey boundary, since the denominator of eq.~\ref{eq:VTFE1} only includes volumes within the survey region. This is a significant advantage over the two methods described above.

However, an important caveat is that the \emph{opposite} problem can and does occur: the necessarily high density of buffer particles placed at the boundaries means that their Voronoi cells encroach within the survey region, at the expense of the `edge' galaxies (Section~\ref{subsection:void-finding}). If this is not corrected for the density does not approach the survey mean at large distances from the void centre (i.e. $\rho(r)>\overline{\rho}$ at large $r$). Fortunately the correction is simple, at least statistically. Since the total survey volume is known and greater than the sum of the Voronoi volumes of all galaxies, the difference can be ascribed entirely to buffer encroachment. We therefore apply a uniform statistical upweighting of the Voronoi volumes of edge galaxies to correct for this.\footnote{One might imagine that the answer is to exclude edge galaxies entirely from the profile determination. However, this introduces a subtle bias of its own, since galaxies in denser environments are less likely to be adjacent to buffer particles. Thus the mean Voronoi volume of non-edge galaxies is smaller than the survey mean, and the same problem reappears.}  

%==================Fig. 4: methods=======================%
\begin{figure}
%\begin{center}
\includegraphics[width=85mm]{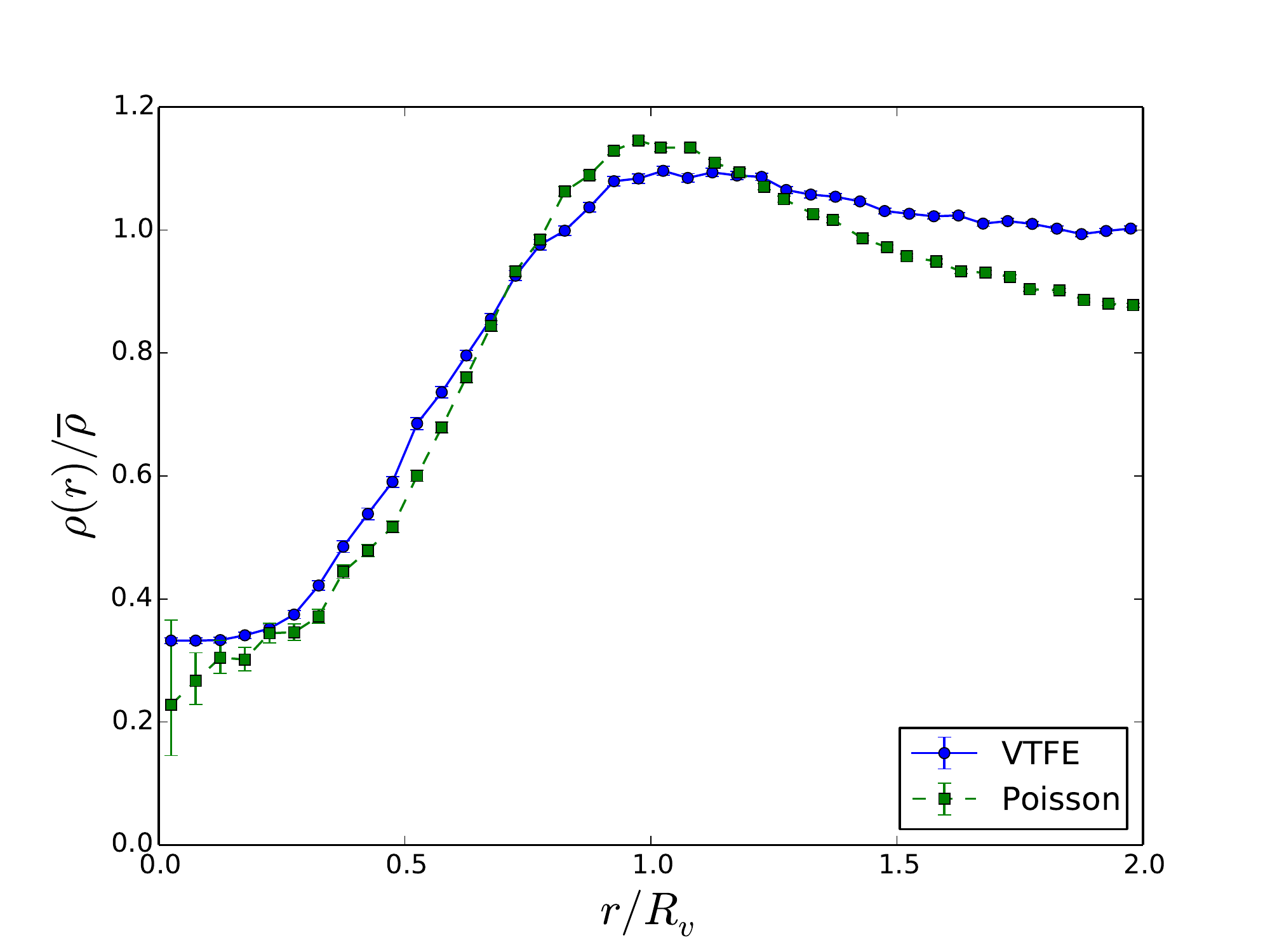}
\caption{Comparison between the stacked void density profiles obtained using the Poisson method (equation~\ref{eq:Poisson1}) and the VTFE method (equation~\ref{eq:VTFE1}), Within $r\lesssim R_v$ the two methods produce similar but not identical results, as discussed in Section~\ref{subsection:comparison}. At larger radii the drop-off in the Poisson-reconstructed profile is an artefact due to leakage effects at the JDim sample boundaries. The VTFE profile, which already corrects for this effect, approaches the mean density at large distances.} 
\label{figure:methods}
%\end{center}
\end{figure}
%==================Fig. 4: methods=======================%

The results we obtain using this method for voids of the three different types are shown in Fig.~\ref{figure:types}. As expected, the differences in the threshold values chosen clearly affect the stacked profile. For instance, Type2 voids, which have a stricter cut on $\rho_\rmn{min}$, are clearly more underdense at the centre than either of the other two classes, and have smaller compensating overdensities at the edges. On the other hand, whereas Type1 and Type3 voids both have the same $\rho_\rmn{min}$ threshold, the stricter restrictions on zone merging applied for Type3 reduce the number of intervening high-density filaments and walls that may lie within the void radius, thereby producing slightly lower densities at all distances from the centre. 

We find that a simple four-parameter fitting formula of the form
\beq
\label{eq:fittingform}
\frac{\rho(r)}{\overline\rho} = 1 +\delta\left(\frac{1-\left(r/r_s\right)^\alpha}{1+\left(r/r_s\right)^\beta}\right)
\eeq
provides a reasonable fit to the simulation data in each case. For the case of Type3 voids, which are the ones we shall primarily focus on, the profile is well described by the values $\delta=-0.69$, $r_s = 0.81R_v$, $\alpha=1.57$ and $\beta=5.72$. Note that whereas the form of equation~\ref{eq:fittingform} is specifically chosen to resemble that used by \cite{Hamaus:2014fma}, the parameter values we find are very different. In particular, unlike the result obtained by those authors, ours is a \emph{single-parameter} rescaling, i.e. that only one parameter in eq.~\ref{eq:fittingform} is dependent on the void size.

Comparing the profiles of Type1 and Type2 voids it is clear that Type1 have both a shallower density minimum in the centre and a a higher overdensity in the compensating shell at $r\sim R_v$. It is worth noting that this is in excellent qualitative agreement with the profiles of the simulated ISW temperature imprints of these voids found by \citet{Hotchkiss:2014}. Since the ISW imprints correspond to the projected gravitational potential along the void direction, this suggests that the number density profiles we show here correspond to similar profiles in the distribution of \emph{mass} at these locations.  

%==================Fig. 5: edgeeffects=======================%
\begin{figure*}
%\begin{center}
\includegraphics[width=170mm]{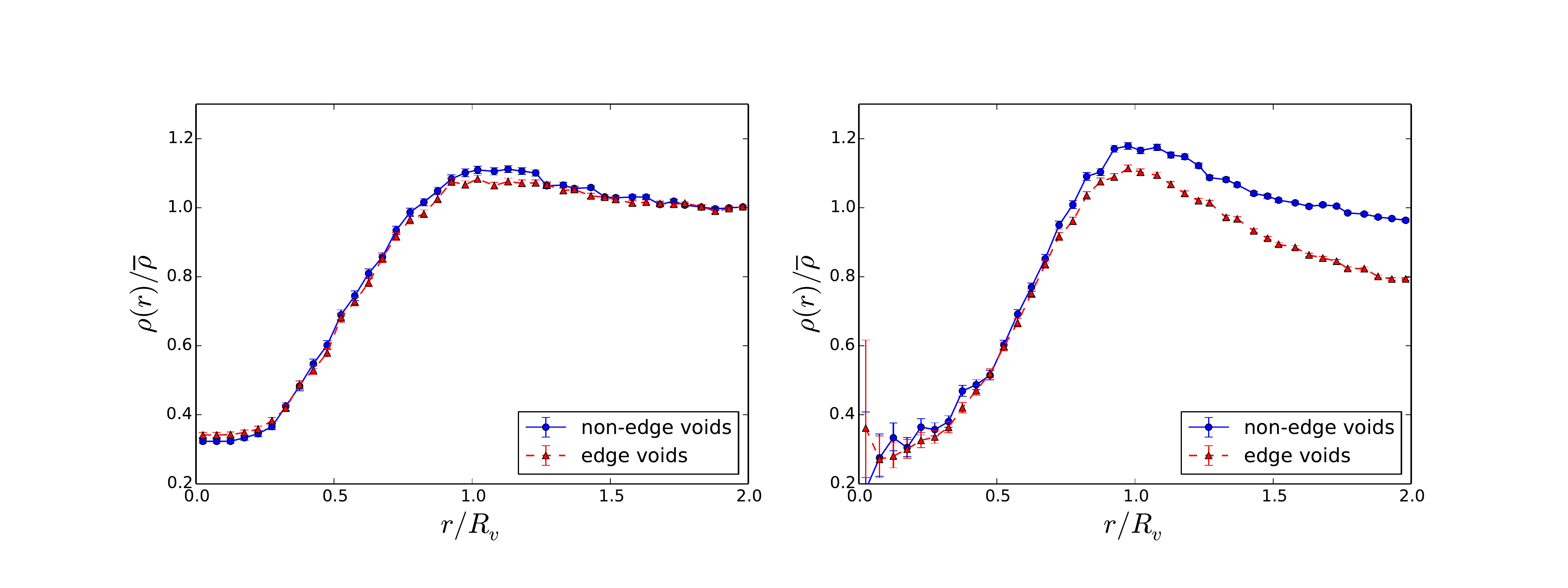}
\caption{The effect of boundary leakage on the void density profile reconstruction. \emph{Left}: Density profiles reconstructed using the VTFE method (equation~\ref{eq:VTFE1}) for voids neighbouring the JDim sample boundaries and those far removed from them show little or no difference as the leakage effect has been corrected for. \emph{Right}: When using the Poisson reconstruction method (equation~\ref{eq:Poisson1}), the two samples of voids give very different density profiles because of leakage effects, showing that this is an inferior method.} 
\label{figure:edgeeffects}
%\end{center}
\end{figure*}
%==================Fig. 5: edgeeffects=======================%

%=========Subsection 4.4: Comparison of Poisson and VTFE methods===========%
\subsection{Comparison of Poisson and VTFE methods}
\label{subsection:comparison}

We now compare the results obtained from the Poisson and VTFE profiling methods. Fig.~\ref{figure:methods} shows the resulting average rescaled profiles out to $2$ times the void radius for the stack of Type3 voids from the JDim sample. The mean radius of this set of voids is $\sim73\;h^{-1}$Mpc although the distribution is significantly skewed towards large radii. We have used a uniform bin spacing of $\Delta = 0.05R_v$.

At radial distances $r\lesssim R_v$, the two profile measurement methods give reasonably similar results, although the Poisson error bars are larger in the void interiors. The VTFE profile appears to be a slightly smoothed version of that obtained by the Poisson method in this region, producing slightly smaller density contrasts both in the void interior and in the surrounding overdense wall. This is due to the fact that the Voronoi tessellation ties the density measurement to the \emph{particle} locations, rather than to the centre of each Voronoi cell. In the interiors of voids (or more generally in the presence of density gradients) tracer particles are slightly shifted away from the cell centres, which produces the smoothing effect.

The more interesting comparison between the two methods is in the region $r>R_v$. Whereas the VTFE profile, after a small overdense compensating wall, reverts to the mean density at large radial distances, the Poisson profile falls \emph{below} the mean. This effect is entirely artificial and is due to the fact that at large distances the spherical shells extend beyond the survey region, which the Poisson method does not account for. This occurs even in our simulated JDim sample, because as noted in Section~\ref{subsection:void-finding}, we apply low and high redshift cuts to mimic a real survey rather than using a cubic simulation box.

To better demonstrate this leakage effect, we split the stack of voids into `edge' voids (in the sense described in Section~\ref{subsection:void-finding}) and those that are far from the redshift caps, and plot the profiles of each group separately. The right panel of Fig.~\ref{figure:edgeeffects} shows that for the Poisson profiling method there is a clear difference between the profiles of voids lying close to or far away from survey edges. As expected, the leakage effect is much more important for voids close to the edges, but it is also present to a lesser extent for non-edge voids. It also starts to become important within the void interior, i.e. at $r\lesssim R_v$. On the other hand, the left panel shows that the VTFE method, where the leakage correction has already been applied, returns the same profile for both subsets.

Clearly this leakage effect is of great importance and must be accounted for when determining void profiles. When using a cubic simulation box with periodic boundary conditions \citep[e.g.][]{Ricciardelli:2014,Hamaus:2014fma}, it does not appear. However, real galaxy survey data are not presented in cubic boxes, and the survey geometry is generally far more complex than the simple low- and high-redshift cuts we have applied to our simulated Jubilee data. Even an ideal full-sky survey will still have significant boundary effects due to masking of foreground stars in our own Galaxy. As a result the majority of voids found in real galaxy surveys do lie close to a survey boundary \citep{Nadathur:2014a}. Any comparison between simulated and observed void density profiles that does not account for this effect is therefore invalid.

It is possible to account for this leakage within the Poisson method, by computing the percentage of each shell in each void that lies within the survey region. For complex survey geometries and given the inevitable presence of holes due to bright stars and other effects, this task will be computationally expensive. On the other hand, applying such a correction in the VTFE method, as we have done here, is straightforward. This is a major reason to prefer the VTFE profile reconstruction method. Other advantages are that the VTFE estimator is less prone to discreteness noise, and that it represents the best available approximation of the density field actually seen by the void-finding algorithm. 

The smoothing effect discussed above means that the density profiles returned by the VTFE estimator may not correspond to the `true' galaxy number density field. However, this distinction does not directly affect other measurable properties of voids, such as their gravitational lensing effect, which depend on the distribution of \emph{mass}. We work solely with number densities rather than mass densities in this paper. Our primary concern is therefore to obtain a measure of the profile that does not depend on survey geometry effects, thus allowing comparison between simulation and real galaxy data. The VTFE estimator satisfies this primary condition. We will now show that it is also independent of the void radius and redshift.

%==============Section 5: Self-similarity======================%

\section{Self-similarity of void density profiles}
\label{section:self-similarity}

The fundamental assumption made in measuring the mean radial profile of a stack of voids is that smaller subsets of the full stack do not have systematically different characteristics. The existence or otherwise of such subsets depends on the selection criteria used to define the stack. Only for samples of voids that are truly \emph{self-similar} is the concept of the mean profile meaningful.

\begin{figure}
%\begin{center}
\includegraphics[width=85mm]{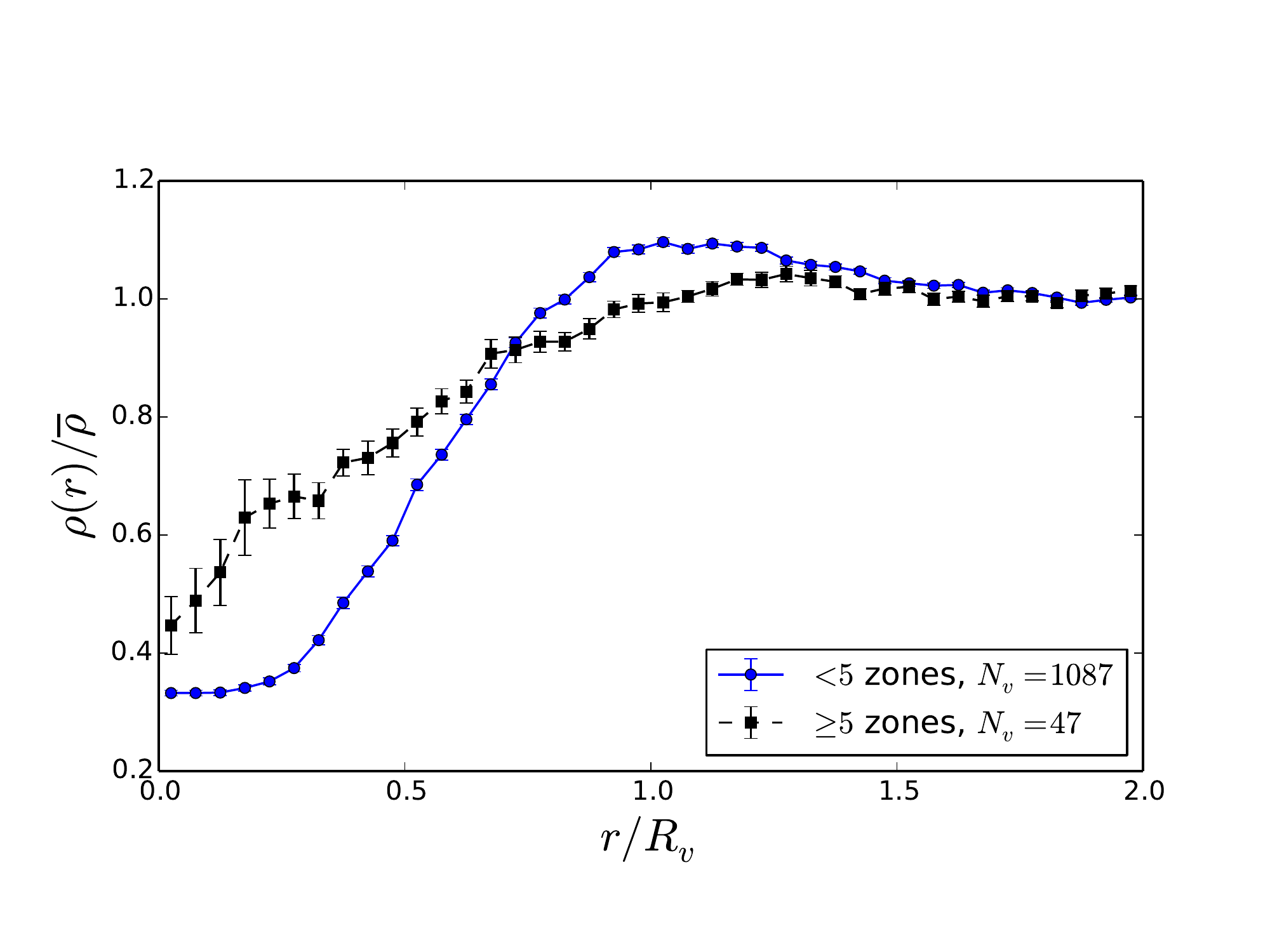}
\caption{Stacked density profiles for the 1087 Type3 voids compared with the equivalent profile for the 47 outlier voids which meet all the selection criteria for Type3 voids except the restriction on the maximum number of merged zones.} 
\label{figure:T3_5z}
%\end{center}
\end{figure}

As an example, in Fig.~\ref{figure:T3_5z} we reproduce the mean stacked profile for the 1087 JDim Type3 voids shown in Fig.~\ref{figure:methods} and compare it to the corresponding profile for the smaller set of 47 voids which meet the same selection cuts on $\rho_\rmn{min}$, $\rho_\rmn{link}$ and $r$, but consist of 5 or more merged local density minima or zones. Clearly there is a large difference between the two reconstructed profiles: this is the reason that we introduced the cut on the maximum number of merged zones in Section~\ref{subsection:void-finding}. Voids formed of a large number of merged zones contain more intervening high-density filaments and walls; in addition the poisition of the void barycentre is more likely to be substantially shifted from the location of the minimum density. These two effects combine to produce the flattening of the profile seen.

%==================Fig. 6: 5z_comp=======================%
%==================Fig. 6: 5z_comp=======================%

%==================Fig. 7: quartiles=======================%
\begin{figure*}
%\begin{center}
\includegraphics[width=170mm]{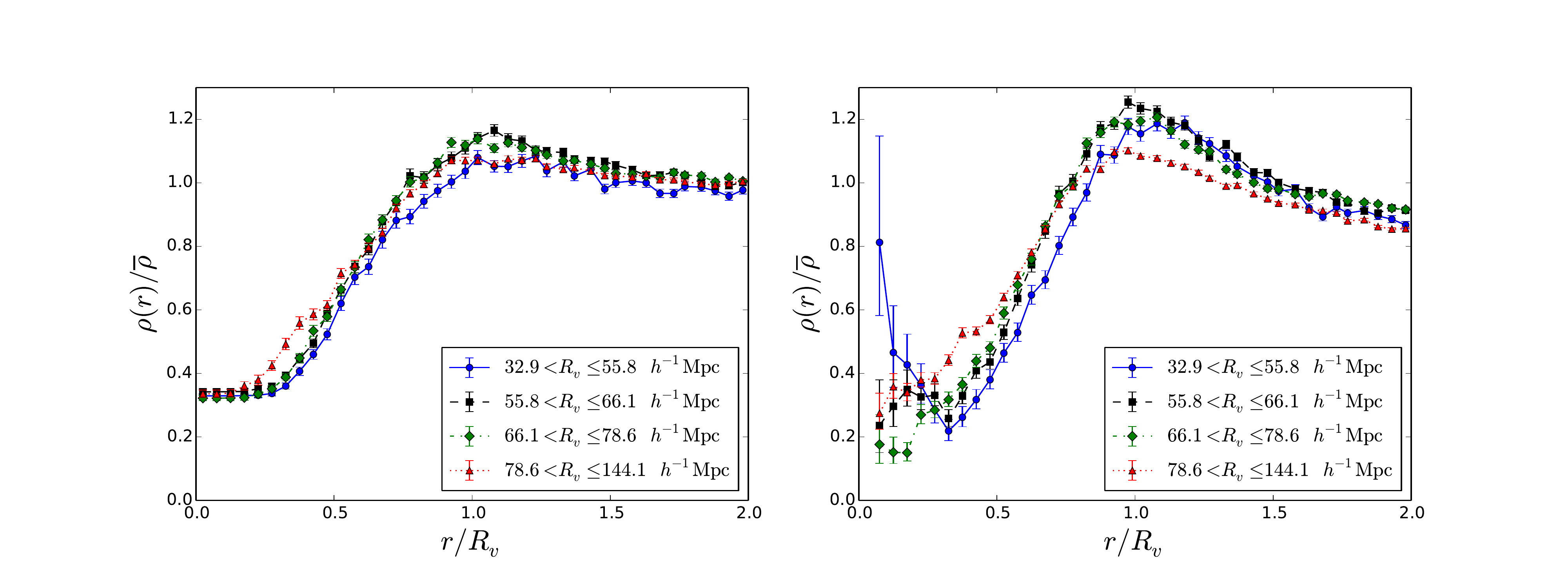}
\caption{Comparison of the mean density profile recovered for voids of different sizes. \emph{Left}: Mean profiles of stacks consisting of different quartiles of Type3 voids ranked by effective radius $R_v$, as reconstructed using the VTFE method. The profiles are very similar and there is no visible trend with void size. \emph{Right}: The same as the left panel, but for profiles reconstructed using the Poisson method.} 
\label{figure:4zquartiles}
%\end{center}
\end{figure*}
%==================Fig. 7: quartiles=======================%

Although we have excluded these outlier multi-zone voids from our stack, other subsets may still exist which spoil the assumption of self-similarity.\footnote{Note that the multi-zone outliers are still reasonable voids, as they certainly correspond to underdensities that are distinguishable from Poisson noise fluctuations. They merely do not correspond to voids which share the same characteristics as those meeting the Type3 definition.} In particular, since radial distances from the centre are rescaled in units of the size of each void, it is pertinent to check that subsets of voids of different sizes give the same result as the full stack. 

To do so we rank the voids by effective radius and split the stack into four separate quartiles. The left panel of Fig.~\ref{figure:4zquartiles} shows the resulting density profiles for these quartiles obtained using the VTFE method for Type3 voids. There are only small differences between the resultant profiles and --- contrary to previous results \citep{Hamaus:2014fma} --- we see no trend with void size. Our selection criteria have resulted in a sample of voids of widely varying sizes that are truly self-similar. This is one of the main results of this paper. 

For good measure, in the right panel we show the same plot obtained using the Poisson method for profile determination. As expected this shows rather noisier behaviour, especially close to the void centres, but no significant trend with void size is observed. However as noted in Section~\ref{subsection:comparison}, the Poisson method \emph{does} give significant differences between `edge' and `non-edge' voids, and is therefore not particularly suitable for this comparison as the relative contribution of edge voids varies across the quartiles. In particular, the largest voids are more likely to be adjacent to boundaries of our simulated region, for obvious reasons. The very smallest voids are also disproportionately likely to be edge voids: their small size is a result of truncation of the watershed merger by the boundary edge.

\begin{figure}
%\begin{center}
\includegraphics[width=85mm]{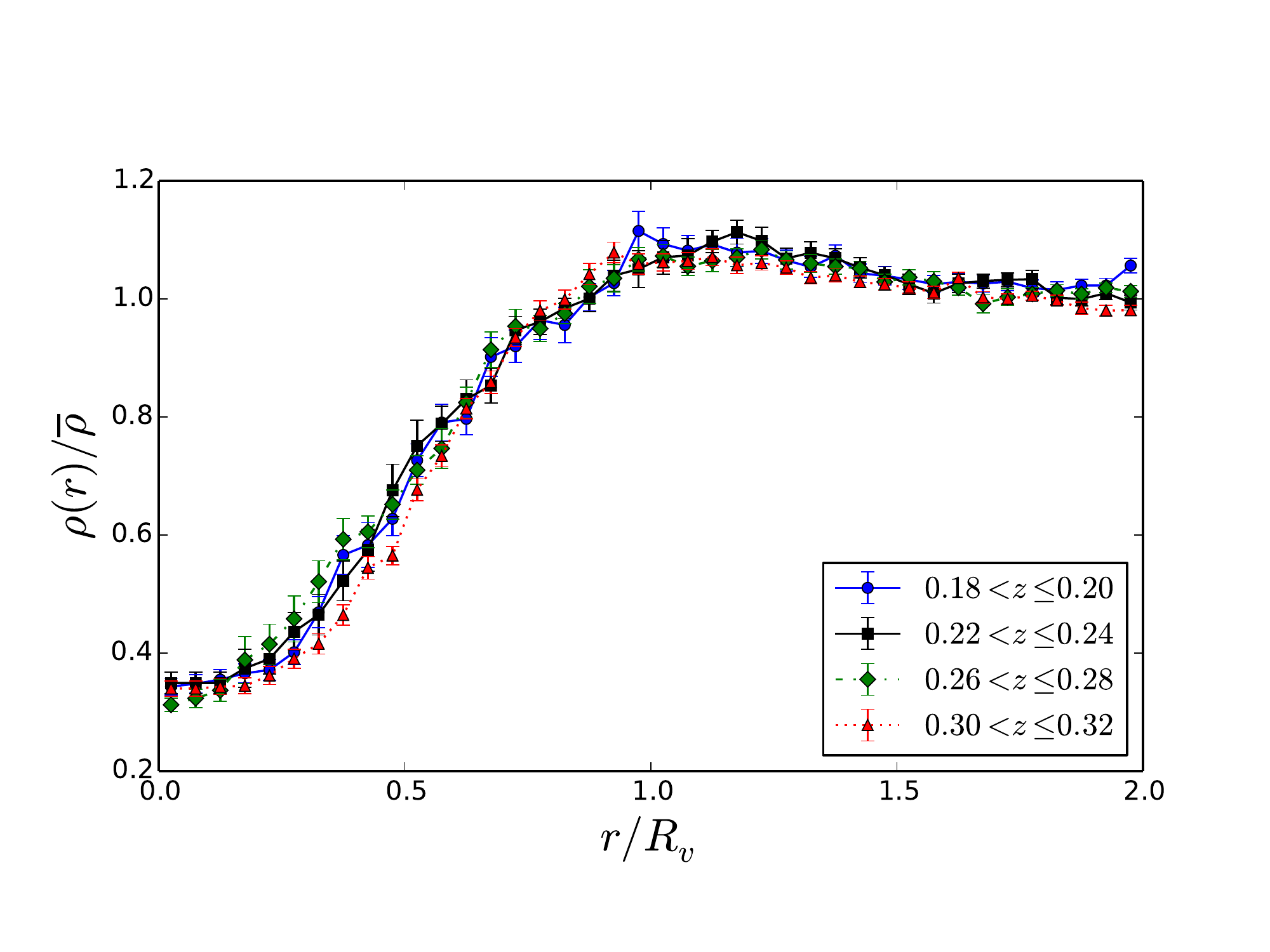}
\caption{Stacked density plots for voids at different redshifts. No difference at all is seen when the redshift is varied over the range available within either of our mock LRG catalogues. Only four sample redshift bins are shown for clarity.} 
\label{figure:redshifts}
%\end{center}
\end{figure}

We also check for a redshift-dependence of the profile by splitting the full stack into subsets at different redshifts. Fig.~\ref{figure:redshifts} shows that, at least within the redshift range probed by our mock LRG catalogues, there is no dependence on redshift at all.

%==================Fig. 8: redshifts=======================%
%==================Fig. 8: redshifts=======================%

We may summarise the results presented in this section as follows. The stacked density profile for a sample of voids depends on the selection criteria used to determine the membership of the sample. However, for reasonable physical selection criteria, chosen primarily in order to distinguish genuine voids from shot noise fluctuations in a discrete point distribution, there is no variation in the mean profile with void size or redshift. This means that our selection criteria applied to the void-finding algorithm select structures that are truly self-similar. Although we have only shown results for Type3 voids from the JDim simulated sample, this self-similarity of void profiles is seen for all void type definitions, and for voids in both of the mock LRG catalogues. 

%==============Section 6: Universality========================%

\section{Universality of void density profiles}
\label{section:universality}

\begin{figure}
%\begin{center}
\includegraphics[width=85mm]{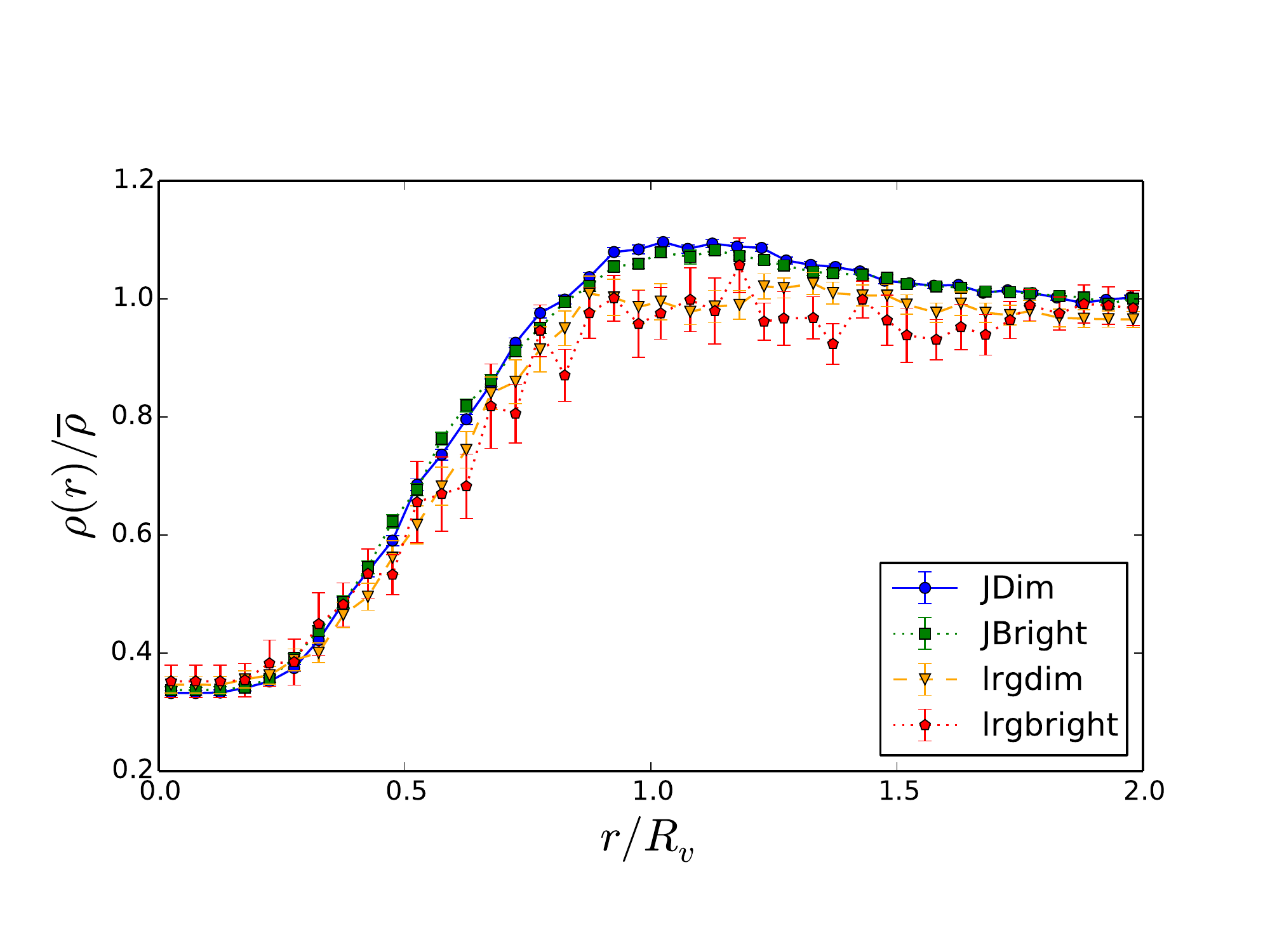}
\caption{Stacked density profiles for simulated and real voids. Real voids are taken from the \emph{lrgdim} and \emph{lrgbright} catalogues, simulated voids from the JDim and JBright mock LRG catalogues designed to match them. All voids are of Type3. Different magnitude cuts for the two LRG samples produce no change in profiles for either simulated or real voids. The real and simulated profiles agree very well with each other, especially in the void interior. } 
\label{figure:Simdatacomp}
%\end{center}
\end{figure}

We now turn to void density profiles seen in SDSS data. Fig.~\ref{figure:Simdatacomp} shows the stacked density profiles for voids found in the \emph{lrgdim} and \emph{lrgbright} samples, compared with those for voids found in their simulation counterparts JDim and JBright. One immediate conclusion is that simulation results match those from real data extremely well, particularly in the interior underdense region of the voids. This match between prediction and observation confirms that our efforts to reproduce the details of dealing with real galaxy data when working with simulation have been largely successful. 

%==================Fig. 9: simvdata=======================%
%==================Fig. 9: simvdata=======================%

%==================Fig. 10: universality=======================%
\begin{figure}
%\begin{center}
\includegraphics[width=85mm]{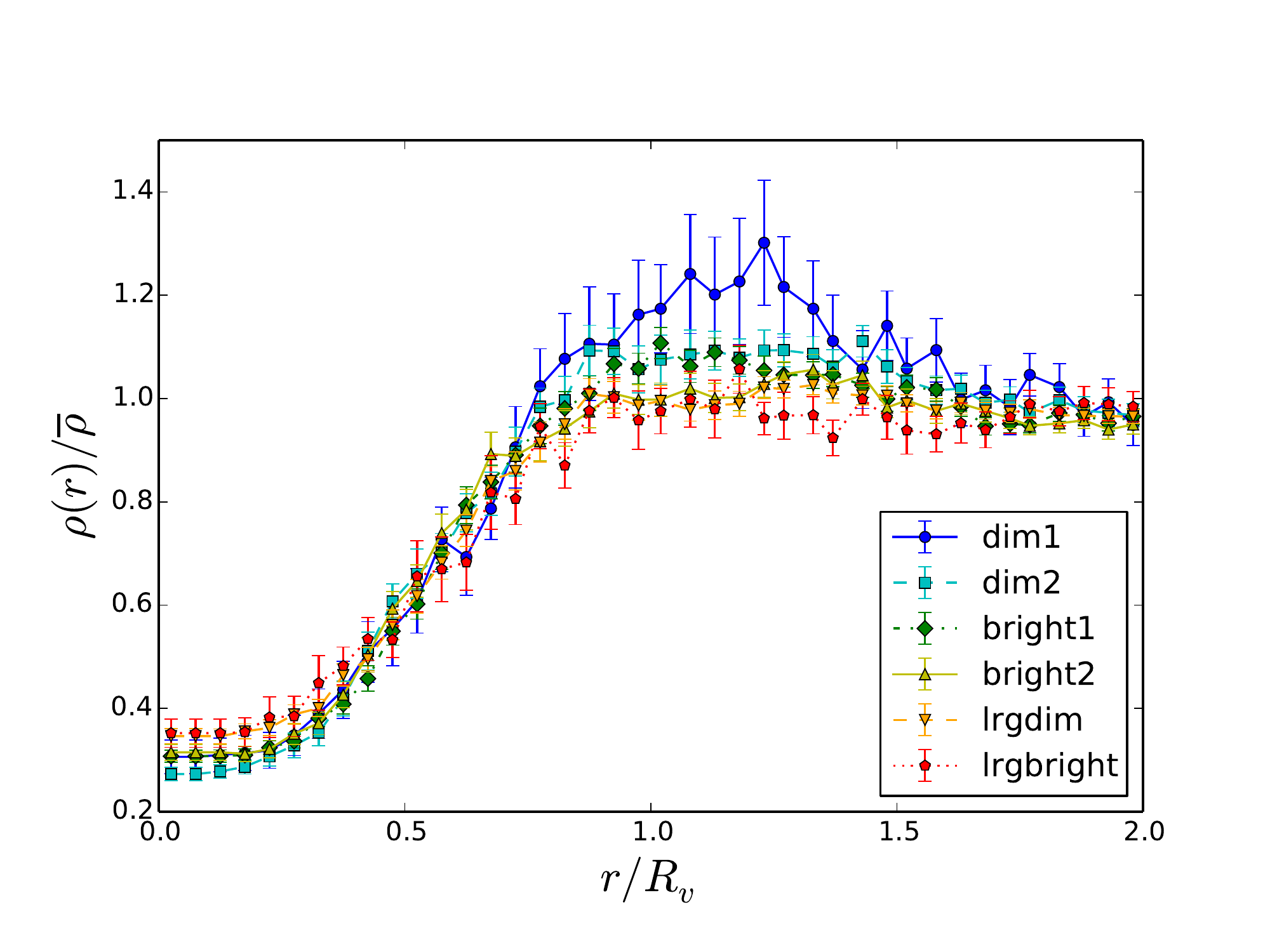}
\caption{Stacked density profiles for voids in the six SDSS galaxy samples \citep{Nadathur:2014a}. All voids are of Type3. Despite wide variation in tracer galaxy luminosity, mean galaxy number density and void size, the profiles show remarkable universality, especially in the interior underdense region.} 
\label{figure:universality}
%\end{center}
\end{figure}
%==================Fig. 10: universality=======================%

There is however a residual small difference between simulated and observed profiles in the region $r\sim R_v$, where in simulation we see an overdense compensating wall at the edges of voids. This may reflect some small inadequacies of the HOD modelling of LRGs in these regions, but a more likely explanation is that these are residual artefacts of the SDSS survey mask. We have shown in Section~\ref{subsection:comparison} that our VTFE method of profile reconstruction does not lead to any systematic differences in profiles between `edge' and `non-edge' voids in simulation. However, the survey edges in our simulation are very simple cuts at the minimum and maximum redshift extents of the mock LRG catalogues, whereas the SDSS mask is highly complex, with many holes. This means that the voids found in real data are often much more severely truncated than those in simulations. This may explain differences in behaviour at large distances from the void centre, for instance if the nearest survey boundary itself lies at the location of the high density ridge. This can be tested by application of the SDSS survey mask to the simulation in order to perfectly mimic the observation, but we leave this task for future work.

The second interesting conclusion from Fig.~\ref{figure:Simdatacomp} is that the change in the magnitude cuts and the consequent difference in sparseness of the two LRG catalogues makes no discernible difference to the mean density profiles of either simulated or real voids. This hints at a degree of universality of the void density profile, independent of the properties of the tracer population. 

To confirm this, we look at the profiles of all Type3 voids from the six different SDSS galaxy samples, which span a wider redshift range (from $z<0.05$ for \emph{dim1} to $0.16<z<0.44$ for \emph{lrgbright}), a wider range of tracer number densities (from $\overline\rho= 2.4\times10^{-2}\;h^{3}\rmn{Mpc}^{-3}$ for \emph{dim1} to $\overline\rho= 2.6\times10^{-5} \;h^{3}\rmn{Mpc}^{-3}$ for \emph{lrgbright}) and a wider range of absolute magnitudes (from $M_r<-18.9$ for \emph{dim1} to $M_g<-21.8$ for \emph{lrgbright}) than those covered in our simulation. Note also that the range of void sizes found in these data also cover a much wider dynamic range than available from the mock LRG catalogues: \emph{dim1} voids have a mean effective radius of $\overline{R_v}=9.6\;h^{-1}$Mpc whereas those from \emph{lrgbright} have $\overline{R_v}=92.8\;h^{-1}$Mpc, almost ten times larger.

In Fig.~\ref{figure:universality} we show the results, confirming the universality of the stacked void profile across the different galaxy samples. This is perhaps the most interesting result of our paper.

Note that the universal nature of profiles is most clear close to the centres of the voids, and profiles from different samples do indeed differ in the high-density wall at the edges of voids. One would expect the properties of galaxy clustering in high density regions to differ markedly between the different galaxy samples simply due to the growth of structure with time, and there does appear to be a trend around $r\sim R_v$ towards greater overdensities at the void edges in the samples at lowest redshift. However we find that these differences do not strongly affect the low-density void interiors.

%==============Section 6: Previous results========================%

\section{Comparison with previous results}
\label{section:previous_results}

Two previous studies have also examined the density profiles of voids found in simulation data \citep{Ricciardelli:2014,Hamaus:2014fma}, but have found results that differ from each other as well as from ours in this paper. \citet{Ricciardelli:2014} agree with our qualitative conclusion that the rescaled stacked profile is independent of void size, but find a quantitatively different profile. There are many possible reasons for this difference, among which are that they use  hydrodynamical as opposed to $N$-body simulations; they trace the density profiles using the dark matter particles and gas in their simulation, whereas we trace the distribution of galaxies; their void-finding algorithm is entirely different; and they use a weighted average scheme for reconstructing the stacked density profile that differs from any of the three estimators discussed here. Given the scale of these differences, even the limited qualitative agreement we find is remarkable.

On the other hand, \citet{Hamaus:2014fma} use a very similar void-finding algorithm to ours, which is also fundamentally based on the {\small ZOBOV} code \citep{Neyrinck:2007gy}, albeit with rather different selection criteria for voids \citep[see also the discussion in][]{NH:2013b,Nadathur:2014a}. Their results are however qualitatively very different --- in particular, they find that smaller voids have significantly lower density interiors than larger voids. 

\begin{figure}
%\begin{center}
\includegraphics[width=85mm]{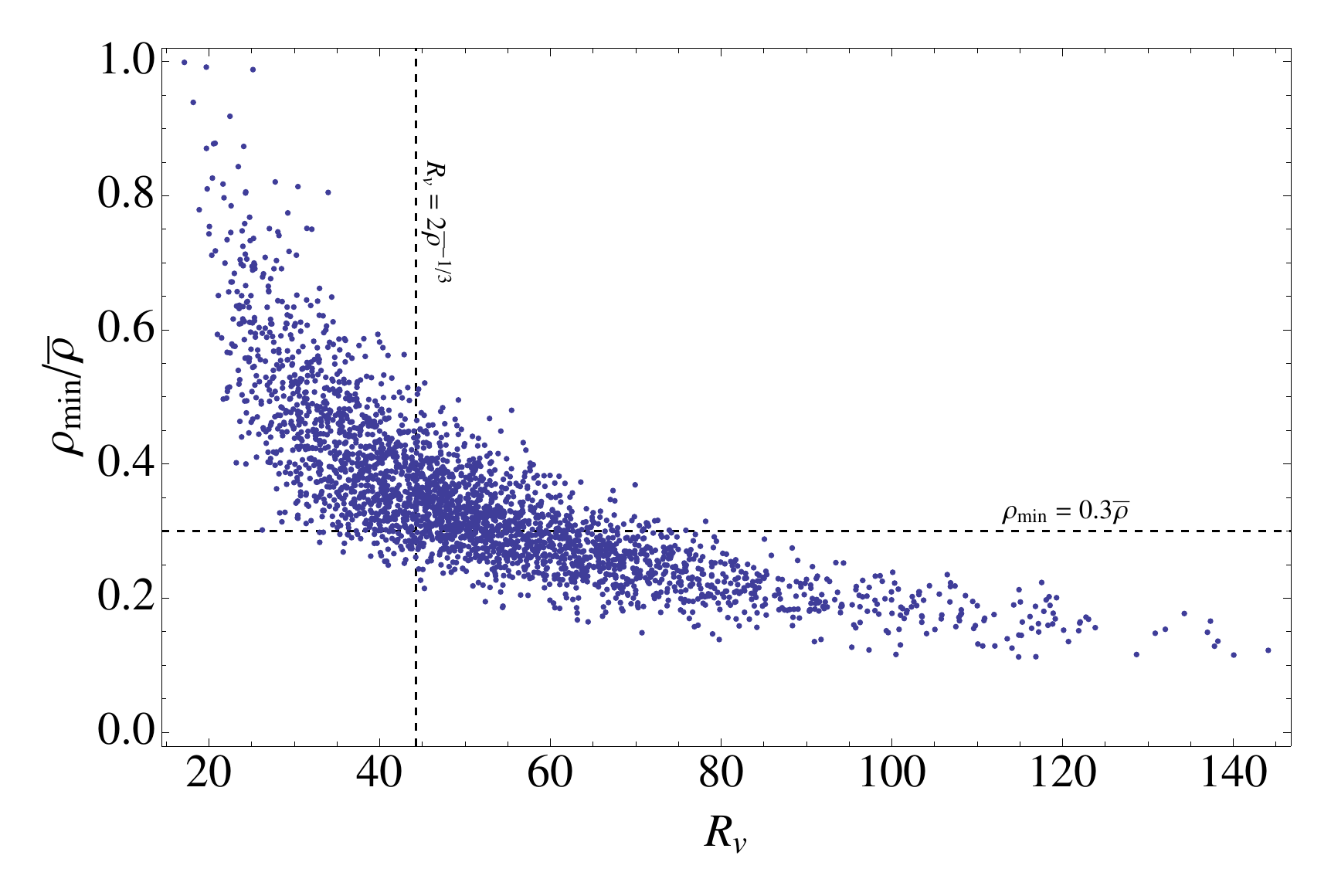}
\caption{Scatter plot showing the relationship between minimum density $\rho_\rmn{min}$ and effective radius $R_v$ for all void candidates in the JDim sample. Thresholds $\rho_\rmn{link}<0.3\overline\rho$ and $r<2$ have been applied to control zone merging. Only those candidate voids lying below the horizontal line at $\rho_\rmn{min}=0.3\overline\rho$ are included in our Type3 sample. The vertical line is at radius value approximately twice the mean galaxy separation: the selection criterion used by \citet{Hamaus:2014fma} corresponds to all candidate voids lying to the right of this line.} 
\label{figure:ZOBOVscatter}
%\end{center}
\end{figure}

On closer inspection this result is very peculiar, as it appears to contradict the known behaviour of the {\small ZOBOV} void-finder. In Fig.~\ref{figure:ZOBOVscatter} we show a scatter plot of the minimum density value $\rho_\rmn{min}$ versus void size $R_v$ for all void candidates (local density minima) from our JDim sample.\footnote{Here we have applied the thresholds $\rho_\rmn{link}<0.3\overline{\rho}$ and $r<2$ to control zone merging as for Type3 voids. However, a different choice of these thresholds does not change the qualitative appearance of this scatter plot.}  The largest void candidates tend to have the deepest underdensities, rather than the other way around. This is a ubiquitous feature of {\small ZOBOV} and the watershed algorithm in general. Changing the properties of the tracer population will alter the mean size of the void candidates and therefore shift the position of the points to the left or right along the $x$-axis, but the overall shape of the plot remains the same  \citep[e.g., see][]{Sutter:2013resp,NH:2013b}.

If anything, we should therefore expect a \emph{negative} correlation between $\rho_\rmn{min}$ and $R_v$, particularly so at the smallest $R_v$ (largest $\rho_\rmn{min}$) values. This can be easily understood in terms of the operation of the watershed algorithm, which merges zones with their deepest neighbour to form voids. This means the deeper the density minimum, the larger the potential for growth and the larger the resulting void.  Note that this negative correlation is also expected if voids are modelled as evolving from isolated spherically symmetric pertrubations of the density field \citep{Sheth:2003py}: for a density field smoothed at a single smoothing scale, the value of which depends on the properties of the tracer population, deeper density minima naturally correspond to larger initial underdense structures \citep{BBKS}. 

%==================Fig. 11: ZOBOV scatter=======================%
%==================Fig. 11: ZOBOV scatter=======================%

This being the case, the question is why \citet{Hamaus:2014fma} appear to see the \emph{opposite} trend in their void profiles. Note that, as shown in Fig.~\ref{figure:ZOBOVscatter}, we apply very different selection criteria to those used in that paper, which is partly responsible for the homogeneity of our sample. If we remove the $\rho_\rmn{min}$ threshold in favour of a minimum radius cut, we do find that self-similarity of the void sample is lost. However, the trend in this case remains in agreement with expectation from Fig.~\ref{figure:ZOBOVscatter} and contrary to that obtained by \citet{Hamaus:2014fma}: the objects with the smallest $R_v$ and the largest $\rho_\rmn{min}$ show the smallest deviations in density from the mean.

A clear difference in simulation approaches is that we find voids in mock galaxy populations, which are explicitly modelled along the past light cone, whereas \citet{Hamaus:2014fma} use randomly sub-sampled dark matter particles from simulation outputs on single time slices as their tracers. This may account for some of the differences in our results if voids in dark matter have very different properties to those in galaxy distributions, although it is not clear why this should be, and \citealt{Sutter:2013ssy} argue against this interpretation. However, if this is the explanation, it should be noted that our result is the one that corresponds to the properties of voids that are actually observable. 

For completeness we note some other possible explanations. One is that the profile reconstruction method used by \citet{Hamaus:2014fma} corresponds to what we have termed the `naive' method here. As we have shown, this method in fact systematically underestimates the densities in precisely those regions where the tracer particle numbers are small (cf. equations~\ref{eq:naive1} and \ref{eq:PoissonExp}), such as the interior of voids. This bias is also worse for smaller voids, which contain fewer tracer particles. Another interesting observation is that the stacked profiles of the largest voids in \citet{Hamaus:2014fma} are very similar to the profile shown in Fig.~\ref{figure:T3_5z} for the stack of voids formed from the merger of 5 or more zones. The number of zones a void is composed of is obviously strongly correlated with its size; it is possible that the trend they see is partly due to this effect. However, whereas we apply very similar thresholds to control zone merging as stated by \citeauthor{Hamaus:2014fma} (in fact, rather looser controls for Type3 voids), we find that only a very small number of outliers are formed of so many zones. 

We cannot be sure which of these effects is the true root cause of the difference in results; it is possible that all of them contribute to some extent, or that some other explanation is required. 

It is worth noting here that the statistical problems affecting the \citet{Hamaus:2014fma} density meaurement in the low-density void interiors will also affect their reconstruction of the velocity profiles. In particular, velocities can only be measured at the locations of the tracer particles in the simulation, but the velocity field is in general non-vanishing even in regions where simulation resolution is not enough to resolve particles, as is inevitably the case for a substantial fraction of shells in the void interiors. This leads to sampling artefacts in the reconstructed volume-weighted velocity field, as also discussed by \citet*{Zhang:2014hra}.

Finally, Fig.~\ref{figure:ZOBOVscatter} also explains the finding \citep{Cai:2013ik} that when no $\rho_\rmn{min}$ cut is applied, the very smallest `voids' found by {\small ZOBOV} are in fact local density minima within large-scale overdensities, and that a cut on $\rho_\rmn{min}$ effectively selects a sample of genuine underdense voids \citep{Hotchkiss:2014}. This is consistent with the findings of other groups \citep[e.g.][]{Ceccarelli:2013,Paz:2013} that the smallest structures returned by other void-finding algorithms also correspond to the `void-in-cloud' scenario \citep{Sheth:2003py}.

%--------------Conclusions------------------%
\section{Discussion}
\label{section:conclusions}

The assumptions of self-similarity and universality of voids and in particular void density profiles have been crucial to several previous studies. Our aim in this work has been to examine the validity of the assumptions. To do so we have used voids identified in mock LRG catalogues on the light cone from the Jubilee simulation and in SDSS galaxy survey data, including both LRG and Main Galaxy samples. Our mock catalogues are designed to be as realistic as possible, and we use exactly the same void identification procedure on simulation and real data. We have shown that standard methods of estimation of the stacked void density profile suffer from systematic bias and volume leakage effects and survey boundaries that affect the comparison between simulation and observation; we therefore use a new estimator based on the Voronoi tessellation density field estimator that accounts for both of these problems.

The selection criteria we use to define a sample of voids are motivated primarily by the need to distinguish genuine voids from random statistical fluctuations, and are dependent on the density minimum $\rho_\rmn{min}$. From simulation results we find that when using these criteria and rescaling distances from the void centres in units of the individual void size, the mean stacked density profile obtained does not depend on the void size or redshift. This means that simulated voids are exactly self-similar objects.

This self-similarity means that within a given set of voids, the density distribution can be characterized by a \emph{single parameter}, the void size, despite the wide range of values of this parameter shown in Fig.~\ref{figure:ZOBOVscatter}. This is in itself a very interesting finding, which may be driven by some unappreciated aspect of the evolution of voids, and is worthy of further study. It also greatly simplifies the theoretical modelling of voids of different sizes and retrospectively justifies studies which have treated all voids by a simple size-based rescaling \citep[e.g.][]{Cai:2013ik,Hotchkiss:2014}. 

However, an important point is that our result of self-similarity is a result of the criteria applied to the selection of objects to classify as `voids'. In particular, the role of the selection cut on $\rho_\rmn{min}$ is crucial: we do \emph{not} find that the self-similarity extends to all void candidates shown in Fig.~\ref{figure:ZOBOVscatter}. We stress again that our default criterion of $\rho_\rmn{min}<0.3\overline\rho$ was chosen to eliminate spurious density minima obtained from Poisson noise, as described by \citet{Nadathur:2014a}.  Cuts applied to other variables do not necessarily achieve the same effect, and we have also confirmed that removing the $\rho_\rmn{min}$ threshold results in a loss of self-similarity primarily due to the inclusion of objects which are not in fact significantly underdense.

Comparison of void profiles from simulated LRG galaxy catalogues and those from SDSS shows a good agreement between our predictions and observation, as shown in Fig.~\ref{figure:Simdatacomp}. This is a vindication of our approach in attempting at every stage to match the procedures applied to our simulated data to those required when dealing with real galaxy surveys. In the past some studies have found differences between simulation data and observation \citep[e.g.][]{Tavasoli:2013,Ricciardelli:2014}, but these studies do not account for the serious impact of survey mask, while also in some cases applying different void-finding procedures to simulation and galaxy data. Our results suggest that when these effects are correctly accounted for these differences disappear.

In addition we have shown that void profiles from SDSS galaxy samples covering a wide range of galaxy magnitudes and number densities display a remarkable degree of universality, being essentially indistinguishable from each other within the void interior (Fig.~\ref{figure:universality}). This greatly extends the results found from simulation and is perhaps our most interesting and significant result.

Note that the mean void sizes in the different stacks in Fig.~\ref{figure:universality} cover a very wide range of values, ranging from $\sim10\;h^{-1}$Mpc to $\sim100\;h^{-1}$Mpc. However, except for the two lowest redshift samples \emph{dim1} and \emph{dim2}, in most cases the scales in question could be decribed as being within the linear regime, i.e. the voids have effective radii of tens of Mpc. This is undoubtedly related to the self-similarity and universality of the profiles seen, and it is no coincidence that the biggest hints of a deviation from the universal profile in Fig.~\ref{figure:universality} are seen for voids in \emph{dim1}.

Throughout this work we have dealt with \emph{number} densities rather than mass densities. This is because whereas numbers of galaxies are simple to count, the relationship to the galaxy mass, let alone the mass of the host halo or the underlying dark matter density, is harder to model. In one sense, of course, the difference is immaterial: the quantity we consider is directly measurable, does not depend on the void properties, and can be compared with observation. Provided alternative models of dark energy or modified gravity predict a measurably different void number density profile, this quantity can be used to obtain cosmological constraints without reference to the mass density. We will explore these issues in future work.

On the other hand, it is of course the \emph{mass} density profile (or more precisely the gravitational potential) which is the relevant quantity in some situations, in particular when studying the gravitational lensing effect of voids. This cannot be directly deduced from the results we have presented here, though we plan to address the issue in future work. However, the existence of a self-similarity in the simulated number density profile strongly suggests that this class of voids can also be described a single average mass density profile. 

\section{Acknowledgements}

We are grateful to Yan-chuan Cai for helpful discussions. The Jubilee simulation was performed on the Juropa supercomputer of the J\"ulich Supercomputing Centre (JSC). SN acknowledges support from Academy of Finland grant 1263714. The research leading to these results has received funding from the European Research Council under the European Union’s Seventh Framework Programme (FP/2007–2013) / ERC Grant Agreement No. [308082]. JMD acknowldeges financial support from AYA2010-21766-C03-01, AYA2012-39475-C02-01 and Consolider-Ingenio 2010 CSD2010-00064. ITI acknowledges support from the Science and Technology Facilities Council (grant number ST/L000652/1) and The Southeast Physics Network (SEPNet).  GY acknowledges support from MINECO (Spain) under research grants AYA2012-31101 and FPA2012-34694.

This research has used data from the SDSS Data Release 7. Funding for the SDSS and SDSS-II has been provided by the Alfred P. Sloan Foundation, the Participating Institutions, the National Science Foundation, the U.S. Department of Energy, the National Aeronautics and Space Administration, the Japanese Monbukagakusho, the Max Planck Society, and the Higher Education Funding Council for England. The SDSS website is \url{http://www.sdss.org/}.

\bibliography{refs.bib}

\providecommand{\noopsort}[1]{}
\begin{thebibliography}{}

\bibitem[\protect\citeauthoryear{Abazajian et~al.,}{Abazajian
  et~al.}{2009}]{Abazajian:2008wr}
Abazajian K.~N.,  et~al., 2009, \apjs, 182, 543

\bibitem[\protect\citeauthoryear{{Aragon-Calvo}, {van de Weygaert},
  {Araya-Melo}, {Platen} \& {Szalay}}{{Aragon-Calvo}
  et~al.}{2010}]{Aragon-Calvo:2010}
{Aragon-Calvo} M.~A.,  {van de Weygaert} R.,  {Araya-Melo} P.~A.,  {Platen} E.,
     {Szalay} A.~S.,  2010, \mnras, 404, L89

\bibitem[\protect\citeauthoryear{Bardeen, Bond, Kaiser \& Szalay}{Bardeen
  et~al.}{1986}]{BBKS}
Bardeen J.~M.,  Bond J.,  Kaiser N.,    Szalay A.,  1986, \apj, 304, 15

\bibitem[\protect\citeauthoryear{{Biswas}, {Alizadeh} \& {Wandelt}}{{Biswas}
  et~al.}{2010}]{Biswas:2010}
{Biswas} R.,  {Alizadeh} E.,    {Wandelt} B.~D.,  2010, \prd, 82, 023002

\bibitem[\protect\citeauthoryear{{Blanton} et~al.,}{{Blanton}
  et~al.}{2005}]{Blanton:2004aa}
{Blanton} M.~R.  et~al., 2005, \aj, 129, 2562

\bibitem[\protect\citeauthoryear{{Bos}, {van de Weygaert}, {Dolag} \&
  {Pettorino}}{{Bos} et~al.}{2012}]{Bos:2012}
{Bos} E.~G.~P.,  {van de Weygaert} R.,  {Dolag} K.,    {Pettorino} V.,  2012,
  \mnras, 426, 440

\bibitem[\protect\citeauthoryear{{Cai}, {Neyrinck}, {Szapudi}, {Cole} \&
  {Frenk}}{{Cai} et~al.}{2013}]{Cai:2013ik}
{Cai} Y.-C.,  {Neyrinck} M.~C.,  {Szapudi} I.,  {Cole} S.,    {Frenk} C.~S.,
  2013, ArXiv e-prints, 1301.6136

\bibitem[\protect\citeauthoryear{{Ceccarelli}, {Paz}, {Lares}, {Padilla} \&
  {Lambas}}{{Ceccarelli} et~al.}{2013}]{Ceccarelli:2013}
{Ceccarelli} L.,  {Paz} D.,  {Lares} M.,  {Padilla} N.,    {Lambas} D.~G.,
  2013, \mnras, 434, 1435

\bibitem[\protect\citeauthoryear{{Clampitt}, {Cai} \& {Li}}{{Clampitt}
  et~al.}{2013}]{Clampitt:2012ub}
{Clampitt} J.,  {Cai} Y.-C.,    {Li} B.,  2013, \mnras, 431, 749

\bibitem[\protect\citeauthoryear{{Clampitt} \& {Jain}}{{Clampitt} \&
  {Jain}}{2014}]{Clampitt:2014}
{Clampitt} J.,  {Jain} B.,  2014, ArXiv e-prints, arXiv:1404.1834

\bibitem[\protect\citeauthoryear{{Colberg} et~al.,}{{Colberg}
  et~al.}{2008}]{Colberg:2008}
{Colberg} J.~M.  et~al., 2008, \mnras, 387, 933

\bibitem[\protect\citeauthoryear{Colberg, Sheth, Diaferio, Gao \&
  Yoshida}{Colberg et~al.}{2005}]{Colberg:2005}
Colberg J.~M.,  Sheth R.~K.,  Diaferio A.,  Gao L.,    Yoshida N.,  2005,
  \mnras, 360, 216

\bibitem[\protect\citeauthoryear{{D'Amico}, {Musso}, {Nore{\~n}a} \&
  {Paranjape}}{{D'Amico} et~al.}{2011}]{D'Amico:2011}
{D'Amico} G.,  {Musso} M.,  {Nore{\~n}a} J.,    {Paranjape} A.,  2011, \prd,
  83, 023521

\bibitem[\protect\citeauthoryear{Eisenstein et~al.,}{Eisenstein
  et~al.}{2005}]{Eisenstein:2005su}
Eisenstein D.~J.,  et~al., 2005, \apj, 633, 560

\bibitem[\protect\citeauthoryear{Flender, Hotchkiss \& Nadathur}{Flender
  et~al.}{2013}]{Flender:2012wu}
Flender S.,  Hotchkiss S.,    Nadathur S.,  2013, JCAP, 1302, 013

\bibitem[\protect\citeauthoryear{Granett, Neyrinck \& Szapudi}{Granett
  et~al.}{2008}]{Granett:2008ju}
Granett B.~R.,  Neyrinck M.~C.,    Szapudi I.,  2008, \apj, 683, L99

\bibitem[\protect\citeauthoryear{{Hamaus}, {Sutter} \& {Wandelt}}{{Hamaus}
  et~al.}{2014}]{Hamaus:2014fma}
{Hamaus} N.,  {Sutter} P.~M.,    {Wandelt} B.~D.,  2014, ArXiv e-prints,
  arXiv:1403.5499

\bibitem[\protect\citeauthoryear{{Hotchkiss}, {Nadathur}, {Gottl{\"o}ber},
  {Iliev}, {Knebe}, {Watson} \& {Yepes}}{{Hotchkiss}
  et~al.}{2014}]{Hotchkiss:2014}
{Hotchkiss} S.,  {Nadathur} S.,  {Gottl{\"o}ber} S.,  {Iliev} I.~T.,  {Knebe}
  A.,  {Watson} W.~A.,    {Yepes} G.,  2014, ArXiv e-prints, arXiv:1405.3552

\bibitem[\protect\citeauthoryear{{Ili{\'c}}, {Langer} \& {Douspis}}{{Ili{\'c}}
  et~al.}{2013}]{Ilic:2013cn}
{Ili{\'c}} S.,  {Langer} M.,    {Douspis} M.,  2013, \aap, 556, A51

\bibitem[\protect\citeauthoryear{{Kamionkowski}, {Verde} \&
  {Jimenez}}{{Kamionkowski} et~al.}{2009}]{Kamionkowski:2009}
{Kamionkowski} M.,  {Verde} L.,    {Jimenez} R.,  2009, \jcap, 1, 10

\bibitem[\protect\citeauthoryear{{Kazin} et~al.,}{{Kazin}
  et~al.}{2010}]{Kazin:2010}
{Kazin} E.~A.  et~al., 2010, \apj, 710, 1444

\bibitem[\protect\citeauthoryear{Komatsu et~al.,}{Komatsu
  et~al.}{2009}]{Komatsu:2008hk}
Komatsu E.,  et~al., 2009, \apjs, 180, 330

\bibitem[\protect\citeauthoryear{{Krause}, {Chang}, {Dor{\'e}} \&
  {Umetsu}}{{Krause} et~al.}{2013}]{Krause:2013}
{Krause} E.,  {Chang} T.-C.,  {Dor{\'e}} O.,    {Umetsu} K.,  2013, \apjl, 762,
  L20

\bibitem[\protect\citeauthoryear{Lacey \& Cole}{Lacey \&
  Cole}{1994}]{Lacey:1994su}
Lacey C.~G.,  Cole S.,  1994, \mnras, 271, 676

\bibitem[\protect\citeauthoryear{Lavaux \& Wandelt}{Lavaux \&
  Wandelt}{2012}]{Lavaux:2011yh}
Lavaux G.,  Wandelt B.~D.,  2012, \apj, 754, 109

\bibitem[\protect\citeauthoryear{Lee \& Park}{Lee \& Park}{2009}]{Lee:2007kq}
Lee J.,  Park D.,  2009, \apj, 696, L10

\bibitem[\protect\citeauthoryear{Li, Zhao \& Koyama}{Li
  et~al.}{2012}]{Li:2011pj}
Li B.,  Zhao G.-B.,    Koyama K.,  2012, \mnras, 421, 3481

\bibitem[\protect\citeauthoryear{{Li} \& {Zhao}}{{Li} \&
  {Zhao}}{2009}]{Li:2009}
{Li} B.,  {Zhao} H.,  2009, \prd, 80, 044027

\bibitem[\protect\citeauthoryear{{Melchior}, {Sutter}, {Sheldon}, {Krause} \&
  {Wandelt}}{{Melchior} et~al.}{2013}]{Melchior:2013}
{Melchior} P.,  {Sutter} P.~M.,  {Sheldon} E.~S.,  {Krause} E.,    {Wandelt}
  B.~D.,  2013, ArXiv e-prints, 1309.2045

\bibitem[\protect\citeauthoryear{{Nadathur} \& {Hotchkiss}}{{Nadathur} \&
  {Hotchkiss}}{2013}]{NH:2013b}
{Nadathur} S.,  {Hotchkiss} S.,  2013, ArXiv e-prints, arXiv:1310.6911

\bibitem[\protect\citeauthoryear{{Nadathur} \& {Hotchkiss}}{{Nadathur} \&
  {Hotchkiss}}{2014}]{Nadathur:2014a}
{Nadathur} S.,  {Hotchkiss} S.,  2014, \mnras, 440, 1248

\bibitem[\protect\citeauthoryear{Nadathur, Hotchkiss \& Sarkar}{Nadathur
  et~al.}{2012}]{Nadathur:2011iu}
Nadathur S.,  Hotchkiss S.,    Sarkar S.,  2012, JCAP, 1206, 042

\bibitem[\protect\citeauthoryear{{Neyrinck}}{{Neyrinck}}{2008}]{Neyrinck:2007gy}
{Neyrinck} M.~C.,  2008, \mnras, 386, 2101

\bibitem[\protect\citeauthoryear{Padmanabhan, Schlegel, Finkbeiner, Barentine,
  Blanton et~al.,}{Padmanabhan et~al.}{2008}]{Padmanabhan:2007zd}
Padmanabhan N.,  Schlegel D.,  Finkbeiner D.,  Barentine J.,  Blanton M.,
  et~al., 2008, \apj, 674, 1217

\bibitem[\protect\citeauthoryear{Pan, Vogeley, Hoyle, Choi \& Park}{Pan
  et~al.}{2012}]{Pan:2011hx}
Pan D.~C.,  Vogeley M.~S.,  Hoyle F.,  Choi Y.-Y.,    Park C.,  2012, \mnras,
  421, 926

\bibitem[\protect\citeauthoryear{{Park} \& {Lee}}{{Park} \&
  {Lee}}{2007}]{Park:2007}
{Park} D.,  {Lee} J.,  2007, \apj, 665, 96

\bibitem[\protect\citeauthoryear{{Paz}, {Lares}, {Ceccarelli}, {Padilla} \&
  {Lambas}}{{Paz} et~al.}{2013}]{Paz:2013}
{Paz} D.,  {Lares} M.,  {Ceccarelli} L.,  {Padilla} N.,    {Lambas} D.~G.,
  2013, \mnras, 436, 3480

\bibitem[\protect\citeauthoryear{{Planck Collaboration} et~al.,}{{Planck
  Collaboration} et~al.}{2013}]{Planck:ISW}
{Planck Collaboration} et~al., 2013, ArXiv e-prints, 1303.5079

\bibitem[\protect\citeauthoryear{Platen, van~de Weygaert \& Jones}{Platen
  et~al.}{2007}]{Platen:2007qk}
Platen E.,  van~de Weygaert R.,    Jones B.~J.,  2007, \mnras, 380, 551

\bibitem[\protect\citeauthoryear{{Platen}, {van de Weygaert}, {Jones}, {Vegter}
  \& {Calvo}}{{Platen} et~al.}{2011}]{Platen:2011}
{Platen} E.,  {van de Weygaert} R.,  {Jones} B.~J.~T.,  {Vegter} G.,    {Calvo}
  M.~A.~A.,  2011, \mnras, 416, 2494

\bibitem[\protect\citeauthoryear{{Ricciardelli}, {Quilis} \&
  {Varela}}{{Ricciardelli} et~al.}{2014}]{Ricciardelli:2014}
{Ricciardelli} E.,  {Quilis} V.,    {Varela} J.,  2014, \mnras, 440, 601

\bibitem[\protect\citeauthoryear{{Ryden}}{{Ryden}}{1995}]{Ryden:1995}
{Ryden} B.~S.,  1995, \apj, 452, 25

\bibitem[\protect\citeauthoryear{{Schaap}}{{Schaap}}{2007}]{Schaap:2007}
{Schaap} W.~E.,  2007, PhD Thesis, University of Gr\"oningen

\bibitem[\protect\citeauthoryear{Sheth \& van~de Weygaert}{Sheth \& van~de
  Weygaert}{2004}]{Sheth:2003py}
Sheth R.~K.,  van~de Weygaert R.,  2004, \mnras, 350, 517

\bibitem[\protect\citeauthoryear{{Sousbie}}{{Sousbie}}{2011}]{Sousbie:2011a}
{Sousbie} T.,  2011, \mnras, 414, 350

\bibitem[\protect\citeauthoryear{Sutter, Lavaux, Wandelt \& Weinberg}{Sutter
  et~al.}{2012}]{Sutter:2012tf}
Sutter P.,  Lavaux G.,  Wandelt B.~D.,    Weinberg D.~H.,  2012, \apj, 761, 187

\bibitem[\protect\citeauthoryear{{Sutter}, {Lavaux}, {Hamaus}, {Wandelt},
  {Weinberg} \& {Warren}}{{Sutter} et~al.}{2014}]{Sutter:2013ssy}
{Sutter} P.~M.,  {Lavaux} G.,  {Hamaus} N.,  {Wandelt} B.~D.,  {Weinberg}
  D.~H.,    {Warren} M.~S.,  2014, \mnras, 442, 462

\bibitem[\protect\citeauthoryear{{Sutter}, {Lavaux}, {Wandelt} \&
  {Weinberg}}{{Sutter} et~al.}{2013}]{Sutter:2013resp}
{Sutter} P.~M.,  {Lavaux} G.,  {Wandelt} B.~D.,    {Weinberg} D.~H.,  2013,
  ArXiv eprints, arXiv:1310.5067

\bibitem[\protect\citeauthoryear{{Tavasoli}, {Vasei} \& {Mohayaee}}{{Tavasoli}
  et~al.}{2013}]{Tavasoli:2013}
{Tavasoli} S.,  {Vasei} K.,    {Mohayaee} R.,  2013, \aap, 553, A15

\bibitem[\protect\citeauthoryear{{van de Weygaert} \& {Schaap}}{{van de
  Weygaert} \& {Schaap}}{2009}]{vdWSchaap:2008}
{van de Weygaert} R.,  {Schaap} W.,  2009, in {Mart{\'{\i}}nez}, V.~J. and
  {Saar}, E. and {Mart{\'{\i}}nez-Gonz{\'a}lez}, E. and {Pons-Border{\'{\i}}a},
  M.-J., eds., Berlin Springer Verlag \emph{Data Analysis in Cosmology}.
  (arXiv:0708.1441)

\bibitem[\protect\citeauthoryear{{Watson}, {\noopsort{a}} {Iliev}, {Diego},
  {Gottl{\"o}ber}, {Knebe}, {Mart{\'{\i}}nez-Gonz{\'a}lez} \& {Yepes}}{{Watson}
  et~al.}{2014}]{Watson:2013mea}
{Watson} W.~A.,  {\noopsort{a}} {Iliev} I.~T.,  {Diego} J.~M.,  {Gottl{\"o}ber}
  S.,  {Knebe} A.,  {Mart{\'{\i}}nez-Gonz{\'a}lez} E.,    {Yepes} G.,  2014,
  \mnras, 437, 3776

\bibitem[\protect\citeauthoryear{{Watson} et~al.,}{{Watson}
  et~al.}{2014}]{Watson:2013cxa}
{Watson} W.~A.  et~al., 2014, \mnras, 438, 412

\bibitem[\protect\citeauthoryear{{Watson}, {Iliev}, {D'Aloisio}, {Knebe},
  {Shapiro} \& {Yepes}}{{Watson} et~al.}{2013}]{Watson:2012mt}
{Watson} W.~A.,  {Iliev} I.~T.,  {D'Aloisio} A.,  {Knebe} A.,  {Shapiro} P.~R.,
     {Yepes} G.,  2013, \mnras, 433, 1230

\bibitem[\protect\citeauthoryear{Wen, Han \& Liu}{Wen
  et~al.}{2012}]{Wen:2012tm}
Wen Z.,  Han J.,    Liu F.,  2012, \apjs, 199, 34

\bibitem[\protect\citeauthoryear{Zeldovich}{Zeldovich}{1970}]{Zeldovich:1969sb}
Zeldovich Y.,  1970, \aap, 5, 84

\bibitem[\protect\citeauthoryear{{Zhang}, {Zheng} \& {Jing}}{{Zhang}
  et~al.}{2014}]{Zhang:2014hra}
{Zhang} P.,  {Zheng} Y.,    {Jing} Y.,  2014, ArXiv e-prints, arXiv:1405.7125

\bibitem[\protect\citeauthoryear{Zheng, Zehavi, Eisenstein, Weinberg \&
  Jing}{Zheng et~al.}{2009}]{Zheng:2008np}
Zheng Z.,  Zehavi I.,  Eisenstein D.~J.,  Weinberg D.~H.,    Jing Y.,  2009,
  \apj, 707, 554

\bibitem[\protect\citeauthoryear{{Zitrin}, {Bartelmann}, {Umetsu}, {Oguri} \&
  {Broadhurst}}{{Zitrin} et~al.}{2012}]{Zitrin:2012}
{Zitrin} A.,  {Bartelmann} M.,  {Umetsu} K.,  {Oguri} M.,    {Broadhurst} T.,
  2012, \mnras, 426, 2944

\end{thebibliography}
\bibliographystyle{mn2e}

\label{lastpage}
\end{document}